%% file: main.tex
\documentclass[11pt]{article}

\include{init}

\title{
\vspace{-2.0cm}
\begin{flushright}
    \small
    CERN-TH-2024-*
\end{flushright}
\vspace{2cm}

Bayesian Inference with Gaussian Processes for the Determination of Parton Distribution Functions}
\author[a]{Alessandro Candido}
\author[b]{Luigi Del Debbio} 
\author[c,d]{Tommaso Giani} 
\author[e]{Giacomo Petrillo}
\affil[a]{CERN, Theoretical Physics Department, CH-1211 Geneva 23, Switzerland}
\affil[b]{Higgs Centre for Theoretical Physics, School of Physics and Astronomy,
Peter~Guthrie~Tait~Road, Edinburgh EH9 3 FD, United Kingdom.}
\affil[c]{Department of Physics and Astronomy, Vrije Universiteit, NL 1081 HV Amsterdam}
\affil[d]{Nikhef Theory Group, Science Park 105, 1098 XG Amsterdam, The Netherlands}
\affil[e]{Dipartimento di Statistica, Informatica, Applicazioni ``Giuseppe Parenti'' (DISIA), Universit\`a di Firenze,
Viale Morgagni 59, 50134 Firenze, Italy}

\date{}
\makeindex

\begin{document}

\maketitle

\begin{abstract}
    We discuss a Bayesian methodology for the solution of the inverse problem
    underlying the determination of parton distribution functions (PDFs).
    In our approach, Gaussian Processes (GPs) are used to model the PDF prior, while Bayes' 
    theorem is used in order to determine the posterior distribution of the PDFs given a set
    of data. 
    We discuss the general formalism, the Bayesian inference at the level of both 
    parameters and hyperparameters, and the simplifications which occur when the observable 
    entering the analysis is linear in the PDF. 
    We benchmark the new methodology in two simple examples for the determination
    of a single PDF flavor from a set of Deep Inelastic Scattering (DIS) data and from a 
    set of equal-time correlators computed using lattice QCD. 
    We discuss our results, showing how the proposed methodology allows for a well-defined 
    statistical interpretation of the different sources of errors entering the PDF uncertainty,
    and how results can be validated a posteriori.
\end{abstract}

\include{intro}

\include{gauss_tg}
\include{example_ns}

\include{closure}

\include{summary}

\appendix

\include{prior_pdf}
\include{kinlim}

\bibliographystyle{UTPstyle}
\bibliography{biblio}

\end{document}

%% file: init.tex
\usepackage{amsmath}
\usepackage{amsfonts}
\usepackage{amssymb}
\usepackage{wasysym}
\usepackage{authblk}
\usepackage{dsfont}
\usepackage{pifont}
\usepackage{booktabs}
\usepackage{tabularx}
\usepackage{siunitx}
\usepackage{graphicx}
\usepackage{epstopdf}
\usepackage{epsfig}
\usepackage{framed}
\usepackage{makeidx}
\usepackage{simplewick}
\usepackage{hyperref}
\usepackage{placeins}
\usepackage{bbold}
\usepackage[font=small,labelfont=bf]{caption}
\usepackage{tikz}
\usetikzlibrary{positioning}
\usetikzlibrary{arrows}
\usetikzlibrary{colorbrewer}

\DeclareSIUnit{\barn}{b}

\newcommand{\sfa}{{\sf a}}

\newcommand{\fk}{\text{FK}}
\newcommand{\mbx}{\mathbf{x}}
\newcommand{\mbm}{\mathbf{m}}
\newcommand{\mbf}{\mathbf{f}}
\newcommand{\covthdat}{C_{YT}}

\newcommand{\ie}{{\it i.e.}}

\newcommand{\viz}{{\it viz.}}

\newcommand{\ndat}{N_{\mathrm{dat}}}

\newcommand{\cov}{\mathrm{Cov}}
\newcommand{\FKtab}{(\mathrm{FK})}
\newcommand{\FKtabT}{(\mathrm{FK})^T}
\newcommand{\GP}{\mathcal{GP}}

\newcommand{\FKtabs}{(\fk^*)}

\newcommand{\pdet}{\operatorname{pdet}}

\graphicspath{{./figures/}}

\let\oldmarginpar\marginpar
\renewcommand{\marginpar}[1]{\oldmarginpar{\sffamily\scriptsize #1}}

%% file: intro.tex
\section{Introduction}
\label{sec:Intro}

The determination of one or more continuous functions knowing a finite set 
of experimental observations is notoriously an ill-posed problem, 
which goes under the name of inverse problem.
The extraction of Parton Distribution Functions (PDFs) from experimental and lattice
data is an example of this in high-energy physics. 
PDFs are an essential input to perform analyses and computations in collider
phenomenology, and are required for a number of precision studies concerning the
determination of standard model parameters and searches for new physics.
PDF determinations currently used for phenomenological studies include
MSHT20~\cite{Bailey:2020ooq}, CTEQ18~\cite{Hou:2019efy}, NNPDF4.0~\cite{NNPDF:2021njg},
HERAPDF2.0~\cite{H1:2015ubc}.
A strong dependence on the PDFs has been observed in recent determinations of the strong
coupling $\alpha_s$ and of the $W$ boson mass carried out by the ATLAS
collaboration~\cite{ATLAS:2023lhg, ATLAS-CONF-2023-004}: when changing the PDF set used
as input in the analysis, the output fluctuates by an amount which is bigger than the
quoted PDF error.
These discrepancies could be generated by some incompatibilities between independent PDF 
determinations and
raises the question whether all the relevant sources of uncertainty are properly
accounted for in the quoted PDF error. Comparing the results obtained with 
different methodologies is one way to test the robustness of the error estimates,
and combination studies have been performed to provide the community with a unique PDF set to 
be used for phenomenology~\cite{PDF4LHCWorkingGroup:2022cjn}.

Given the ill-posed nature of the inverse problem underlying the determination of PDFs,
a regularization method is necessary in order to make the problem well defined.
The regularization reduces the problem to a finite dimensional and solvable one, but it inevitably
introduces some bias, which depends on the specific methodological choices.
The solution to an inverse problem will therefore come with an error associated with the
methodology, which has to be quantified, just like the other uncertainties entering a
PDF fit (uncertainties of the experimental data and of the input standard model
parameters, theory errors due to missing QCD higher orders).
Despite specific efforts in this direction have already been pursued - for example in
Refs.~\cite{Ball_2015,DelDebbio:2021whr} the different components of PDF uncertainty are
qualitatively assessed using the formalism of closure tests - a way to assess
quantitatively the size of the methodological error in a PDF determination is still
 missing.

In this paper we investigate a Bayesian approach to the solution of 
inverse problems, by extending the preliminary work discussed in
Ref.~\cite{Candido:2023nnb}, and further developing some of the ideas introduced in
Ref.~\cite{DelDebbio:2021whr}. The goal is to develop a methodology, orthogonal to those currently used within the
collinear PDF community, where all the relevant sources of uncertainty, including the
methodological one, have a clear mathematical definition. We argue that such methodology
would simplify the discussion around discrepancies of the kind observed, for instance,
in Refs.~\cite{ATLAS:2023lhg, ATLAS-CONF-2023-004}, providing a quantitative estimate of
the different sources of error entering the PDF uncertainty. 

In the fitting methodologies currently used for PDF determinations, the unknown model 
is parameterized in terms of a finite (albeit large) set of parameters, 
which are then fitted to the observed data.
In the Gaussian Processes approach, rather than starting by a parameterization, a
prior probability distribution is introduced for the target model in its original space,
encoding our a priori theoretical knowledge of the unknown target function.
Using Bayes’ theorem, it is possible to determine the posterior distribution of the
solution after taking into account a set of experimental observations. 
This approach has multiple advantages: the inverse problem is well-defined, all the
assumptions made on the model are explicitly stated in the choice of the prior and the
results are given in terms of posterior probability distributions, making all the
relevant uncertainties well-defined from a mathematical point of view. 
On the other hand, as with any other regularization method, the Bayesian approach 
introduces a bias through the choice of a specific prior; the posterior probability
distribution does depend on the choice of the prior and this dependence needs to be
studied and properly quantified. In this paper we will argue that the quantification of 
the existing bias and the different sources of error affecting the final
result is particularly clear in a Bayesian approach. 

Our Bayesian approach relies on promoting the values of the PDFs to stochastic variables, 
whose probability distributions are constrained by experimental data. These posterior probability 
distributions encode all the information about the PDFs. 
A possible way to do this is by using the formalism of Gaussian Processes
(GPs)~\cite{books/lib/RasmussenW06}, through which a suitable prior for the unknown PDFs
can be defined, in terms of a reduced number of hyperparameters. 
GPs have already been used to solve inverse problems in various fields in physics,
from geophysics~\cite{10.1093/gji/ggz520} to lattice QCD~\cite{Hansen:2019idp, Horak:2021syv,Karpie:2019eiq,Horak:2023xfb}.
The main focus of this paper is the study of GPs in the context of PDF determinations,
including the choice of the most suitable kernel (which defines the prior distribution),
the optimization of the corresponding hyperparameters and the way in which 
theoretical knowledge about PDFs -- such as sum rules, kinetic limit, and integrability
constraints -- can be encoded in the prior.

In Sec.~\ref{sec:Gauss} we recall the definition and some well-known properties of GPs,
we set the notation and spell out the different steps of the proposed methodology for
PDF determination. We focus on the case of observables linear in the PDF and we briefly 
discuss what changes are required when quadratic observables are included in the analysis.
In Sec.~\ref{sec:examples} we discuss the choice of a prior distribution for PDFs and we
provide two simple examples concerning the determination of a single PDF flavor from a
set of Deep Inelastic Scattering (DIS) data and lattice equal time correlators.
In Sec.~\ref{sec:discussion} we discuss the results, the quantitative evaluation of the
different sources of uncertainties entering the analysis, and possible ways to validate
the results a posteriori.
Conclusions and outlook are presented in Sec.~\ref{sec:conclusions}.

%% file: gauss_tg.tex
\section{Gaussian Processes for Inference}
\label{sec:Gauss}

In the following, we recall the definition of a Gaussian Process, setting the
notation for the subsequent sections.
Moreover, we describe indetail the case of a GP regression in the presence of
data that depend linearly on the GP, subject to a hyperparameterized prior.
While the case of linearly dependent data is sufficient for the investigation presented 
in this work, we also introduce
a more general case, which allows us to clarify the simplifications observed in
our current study,
and sets the framework for further developments towards a global PDF determination.
Consequently, we are not going to provide an exhaustive presentation about GPs, for
which the reader could refer to the existing literature, such as Ref.~\cite{books/lib/RasmussenW06}.

\subsection{Notation}
\label{sec:GPNotation}

In a Bayesian approach, the true value $f(x)$ of the PDF for each $x\in [0,1]$ is
treated as a random variable.
We should therefore think of $x$ as a continuous index, which parametrizes the elements
of a stochastic process. 
A Gaussian process,
\begin{equation}
  \label{eq:GPDef}
  f \sim \mathcal{GP}\left(m, k\right)\, ,
\end{equation}
is a particular type of stochastic process, whose probability distribution is entirely
specified by two functions, the mean $m(x)$ and the kernel $k(x,x')$.
The values of the function $f$ at any discrete set of points, \[\mathbf{x}=\{x_i; i=1,
\ldots, N\}\, ,\] define a vector of stochastic variables
\begin{align}
    \label{eq:FunVect}
    \mathbf{f} = f(\mathbf{x}) = 
    \begin{pmatrix}
        f_1 \\
        \vdots \\
        f_N
    \end{pmatrix} \in \mathbb{R}^N\,,
    \quad f_i = f(x_i)\,, \quad i=1,\ldots, N\, .
\end{align}
The probability distribution of these variables is an $N$-dimensional Gaussian
distribution, 
\begin{align}
    \label{eq:FunMultiGauss}
    \mathbf{f} \sim \mathcal{N}(\mathbf{m},K)\, ,
\end{align}
whose mean and covariance are given by
\begin{align}
    \label{eq:FunMultiGaussDetails}
    \mathbf{m} &= m(\mathbf{x})\, , \quad
    K = k(\mathbf{x},\mathbf{x}^T)\, ,
\end{align}
and therefore
\begin{align}
    \label{eq:MultiGaussAvgCov}
    E[f_i] &= m_i = m(x_i)\, , \\
    \mathrm{Cov}[f_i,f_j] &= K_{ij} = k(x_i,x_j)\, .
\end{align}

In the following, we will distinguish the points in $x$ for which the value $f(x)$ is 
included in the theoretical prediction for the measurements, and those 
where we want to infer the value of the function.
We denote the former by $\mathbf{x}$ and the latter by $\mathbf{x^*}$. 
The corresponding vectors $\mathbf{f}$ and $\mathbf{f^*}$ are defined as in
Eq.~\eqref{eq:FunVect}. 
In a Bayesian formalism, we define a prior joint distribution for
$(\mathbf{f},\mathbf{f^*})$ and a likelihood function, which will depend on
$\mathbf{f}$.
We can then compute the posterior distribution for $\mathbf{f^*}$ applying Bayes'
theorem. Assuming that 
\[
    \mathbf{f} \in \mathbb{R}^N\, , \quad \mathbf{f^*} \in \mathbb{R}^M\, ,    
\]
then the Gaussian process defined in Eq.~\eqref{eq:GPDef} yields a prior
distribution, 
\begin{align}
    \label{eq:PriorDistr}
    p(\mbf,\mbf^*|\theta) = 
    \frac{1}{\sqrt{\det\left(2\pi K\right)}}
    \exp\left\{
        - \frac12 \left((\mathbf{f}-\mathbf{m})^T,
        (\mathbf{f^*}-\mathbf{m^*})^T\right)
        K^{-1} 
        \begin{pmatrix}
            \mathbf{f}-\mathbf{m} \\
            \mathbf{f^*}-\mathbf{m^*}
        \end{pmatrix}
    \right\}\, ,  
\end{align}
where $K$ is now an $(N+M)\times(N+M)$ matrix~\footnote{We have slightly changed the
notation here, compared to the one we used in Ref.~\cite{Candido:2023nnb}.}, 
\begin{align}
    \label{eq:KFFstar}
    K = \begin{pmatrix}
            k(\mathbf{x},\mathbf{x}^T) & k(\mathbf{x},\mathbf{x^*}^T) \\
            k(\mathbf{x^*},\mathbf{x}^T) & k(\mathbf{x^*},\mathbf{x^*}^T) 
        \end{pmatrix}
      = \begin{pmatrix}
          K_{\mbx\mbx} & K_{\mbx\mbx^*} \\
          K_{\mbx^*\mbx} & K_{\mbx^*\mbx^*}
        \end{pmatrix}\, .
\end{align}
The mean and kernel functions might depend on a set 
of additional parameters, usually referred to as hyperparameters, and 
collectively denoted as $\theta$. The dependence of the prior on the 
hyperparameters is marked explicitly in Eq.~\eqref{eq:PriorDistr}.

\subsection{Data and theory predictions}
\label{sec:FKtab}
In Ref.~\cite{Candido:2023nnb} we distinguished two different types of input: direct
observations of the stochastic process, which we called {\em point-wise} data, and
indirect ones, in which only some functions of the process are actually observed.
In this work we will focus on the more general case of indirect observation.
In particular, all the results are obtained for a likelihood model in which the data
appear as a linear functional of $f$. Sec.~\ref{sec:QuadCase} describes how to go beyond
this assumption.

We denote by $T_I$ the prediction for the $I$-th datapoint, that will be computed as
\begin{equation}
  \label{eq:LinThDat}
  T_I = \int dx\, c_I(x) f(x)\, ,
\end{equation}
where $c_I(x)$ are known functions\footnote{
  The method exposed in the following works for generic linear functionals, including
  those that could not be expressed as integrals of regular functions $c_I(x)$.
  This is the case of the observables analysed in PDF fits, but we limited to this
  form to simplify the presentation.
}.
The $T_I$ are distributed according to a Gaussian, with mean value
and covariance
\begin{align}
  \label{eq:MeanThPred}
  E[T_I] &= \int dx\, c_I(x) m(x)\, , \\
  \label{eq:CovThPred}
  \cov[T_I,T_J] &= \int dx' dx''\, c_I(x') k(x',x'') c_J(x'') = A_{IJ}\,.
\end{align}
In practice, we are going to be interested in cases where the integral above
is computed on a grid of points,
\begin{equation}
  \label{eq:LinThGrid}
  T_I = \sum_{i=1}^{N} \FKtab_{Ii} f_i\, ,
\end{equation}
where $\displaystyle \FKtab_{Ii} = \int c_I(x) p_i(x)$, with $p_i(x)$ an interpolation
polynomial, relative to $x_i$. 

The matrix $\FKtab$ is called an {\em FK-table}\ in the NNPDF jargon, and the notation
reflects this convention.
Note that the case of point-wise data (direct observation) is obtained in this framework
by setting $\FKtab$ to the identity. 
The average and the covariance of the theoretical prediction $T$
induced by the prior probability distribution of $\mathbf{f}$ are given by the
discretized versions of Eqs.~\eqref{eq:MeanThPred} and~\eqref{eq:CovThPred},
\begin{align}
  \label{eq:MeanThDiscr}
  E[T_I] &= \FKtab_{Ij} m_j\, ,\\
  \label{eq:CovThDiscr}
  \cov[T_I,T_J] &= \FKtab_{Ii} \left(K_{\mbx\mbx}\right)_{ij} \FKtabT_{jJ}\, .
\end{align}
The experimental central value for the data point corresponding to $T_I$ is denoted $y_I$. 
Note that $T_I$ is a stochastic variable, while $y_I$ is a constant.
In our model, the likelihood is also assumed to be a multivariate Gaussian distribution,
and the fluctuations of the data around their central values are described by the
\textit{experimental} covariance matrix $C_Y$. 

In the rest of the paper, we will omit the indices like $i,j$ and $I,J$ in the
equations above. Boldface vectors, like  $\mbf$ for instance, refer to vectors
computed by evaluating the function $f$ on a grid of points. Vectors in the
space of data will be denoted by ordinary latin characters; the context should
make it easy to identify these vectors in data space, even though we do not have
any typographic convention to identify them. 

\subsection{Inference for the Model}
\label{sec:GPModel}
Following the discussion in
Ref.~\cite{Candido:2023nnb}, we incorporate the knowledge of linear data by
introducing the stochastic variable
\begin{equation}
  \label{eq:EpsDistr}
  \mathbf{\epsilon} \sim \mathcal{N}\left(0, C_Y\right)\, ,
\end{equation}
and imposing that
\begin{equation}
  \label{eq:DataWithErr}
  \FKtab \mathbf{f} + \mathbf{\epsilon} = y\, ,
\end{equation}
where $y$ are the observed experimental central values and $C_Y$ is the covariance matrix
of the data. The linear dependence of $y$ on $\mathbf{f}$ is
encoded in the matrix $\FKtab$. 

We are interested in the probability distribution of the vector $\mbf$ 
and hyperparameters $\theta$, conditioned on Eq.~\eqref{eq:DataWithErr}, which 
we denote as
\begin{align}
  \label{eq:FullPosterior}
  p\left( \mbf, \theta |  y \right) = 
  p\left( \mbf| \theta, y \right) p\left( \theta | y \right)\,.
\end{align}
The two factors on the right-hand side of the equation are best analysed separately,
since being able to sample both of them is enough to sample the left-hand side. 
We focus here on the first term, while the second factor will be discussed in the
following subsection. 
The function $p\left(\mbf| \theta, y \right)$ denotes the posterior probability distribution 
of the vector $\mbf$ for fixed values of the data and of the hyperparameters $\theta$.
In order to compute it, we note that at the level of prior distributions, the vectors $\mbf$ and $\mbf^*$, 
and the data measurement error $\epsilon$ must be uncorrelated, hence the covariance matrix
describing the joint prior distribution of the three sets of stochastic variables,
$(\mbf,\mbf^*,\mathbf{\epsilon})$, is a block-diagonal
$(N+M+\ndat)\times (N+M+\ndat)$ matrix
\begin{equation}
  \label{eq:ThreeVarsCov}
  \cov = \begin{pmatrix}
    K & 0 \\
    0 & C_Y
  \end{pmatrix}\, ,
\end{equation}
where $K$ is the $(N+M)\times (N+M)$ matrix introduced in
Eq.~\eqref{eq:KFFstar}. Therefore the joint prior is 
\begin{align}
  p(\mbf,\mbf^*,\epsilon|\theta) = 
  &\frac{1}{\sqrt{\det\left(2\pi K\right)}}
  \exp\left\{
    - \frac12 \left((\mathbf{f}-\mathbf{m})^T,
    (\mathbf{f^*}-\mathbf{m^*})^T\right)
    K^{-1} 
    \begin{pmatrix}
        \mathbf{f}-\mathbf{m} \\
        \mathbf{f^*}-\mathbf{m^*} 
    \end{pmatrix}
\right\}
   \nonumber \\
  \label{eq:PriorEpsilonDistr}
  &\quad \times \frac{1}{\sqrt{\det\left(2\pi C_Y\right)}}\, 
  \exp\left\{
    - \frac12 \epsilon^T C_Y^{-1} \epsilon
  \right\}\, .
\end{align}
Conditioning on the observed values $y$ in
Eq.~\eqref{eq:DataWithErr}, 
\begin{align}
  p(\mbf,\mbf^*|\theta,y) \propto 
    & \int d\epsilon\, p(\mbf,\mbf^*,\epsilon|\theta)\, 
      \delta(\FKtab \mathbf{f} + \mathbf{\epsilon} - y) \\
      \propto & 
    \exp\left\{
      - \frac12 \left((\mathbf{f}-\mathbf{m})^T,
      (\mathbf{f^*}-\mathbf{m^*})^T\right)
      K^{-1} 
      \begin{pmatrix}
          \mathbf{f}-\mathbf{m} \\
          \mathbf{f^*}-\mathbf{m^*} 
      \end{pmatrix}
  \right\}
     \nonumber \\
    &\times \, 
    \exp\left\{
      - \frac12 (\FKtab \mathbf{f}  - y)^T 
        C_Y^{-1} (\FKtab \mathbf{f} - y)
    \right\}\, .
\end{align}
The final step to get $p\left(\mbf| \theta,y\right)$ involves marginalizing $\mbf^*$, 
which is readily done remembering 
that $(\mbf,\mbf^*)$ obey a multi-dimensional Gaussian distribution,
\begin{align}
    \int &d\mbf^*\, p(\mbf,\mbf^*|\theta,y) 
    \propto  
    \exp\left\{
      - \frac12 (\mathbf{f}-\mathbf{m})^T
      K_{\mbx\mbx}^{-1} 
          (\mathbf{f}-\mathbf{m})
      \right\} \nonumber \\
    &\quad\quad\quad\times \, 
    \exp\left\{
      - \frac12 (\FKtab \mathbf{f} - y)^T 
        C_Y^{-1} (\FKtab \mathbf{f} - y)
      \right\} \nonumber
      \\
    = & \exp\left\{-S(\mbf; \theta,y)\right\}\,,
\end{align}
so that
\begin{align}
  \label{eq:PosteriorF}
  p(\mbf|\theta,y) = 
  \frac{\exp\left\{-S(\mbf; \theta,y)\right\}}
  {\int d\mbf \,\exp\left\{-S(\mbf; \theta,y)\right\}}\,.
\end{align}
This result was already derived in Eq.~(45) in Ref.~\cite{DelDebbio:2021whr}. 
Note that 
\begin{align}
  \label{eq:LogLossIntroduced}
  S(\mbf; \theta, y) = 
    \frac12 \left\{
      (\mathbf{f}-\mathbf{m})^T
      K_{\mbx\mbx}^{-1} 
          (\mathbf{f}-\mathbf{m}) +
          (\FKtab \mathbf{f} - y)^T 
          C_Y^{-1} (\FKtab \mathbf{f} - y)
    \right\}
\end{align}
is a quadratic form in $\mbf$, therefore the normalization in Eq.~\eqref{eq:PosteriorF}
can be computed analytically, yielding a Gaussian posterior for $\mbf$,
\begin{equation}
  \label{eq:PostDataGauss}
  p\left(\mbf| \theta,y\right)
  = \mathcal{N}\left(\mbf; \mathbf{\tilde m}, \tilde{K}_{\mbx\mbx}\right)\,.
\end{equation}
Its mean $\tilde{m}$ and covariance $\tilde{K}_{\mbx\mbx}$ are given by \footnote{
  See \cite[ex.~7.4, p.~295]{schott2017} for the proof.
}
\begin{align}
  \label{eq:MeanPostData}
  \mathbf{\tilde m} 
    &= \mathbf{m} + K_{\mbx\mbx} \FKtabT\, 
    \covthdat^{+}\,
    \left(\mathbf{y} - \FKtab \mathbf{m}\right)\, ,\\
  \label{eq:CovPostDataBis}
  \tilde{K}_{\mbx\mbx} 
    &= K_{\mbx\mbx} - 
    K_{\mbx\mbx} \FKtabT 
    \covthdat^{+}\, 
    \FKtab K_{\mbx\mbx}\, ,
\end{align}
where we introduced  
\begin{equation}
  \label{eq:covthdatDef}
  \covthdat = \FKtab K_{\mbx\mbx} \FKtabT + C_Y\, ,
\end{equation}
which is the covariance of the vector $\FKtab\, \mathbf{f} + \epsilon$,
and the superscript ``$+$'' denotes the matrix pseudoinverse. 
In the following, we replace the pseudoinverse with the inverse,
and the formulae derived implicitly assume that the corresponding matrices are invertible. 
Eq.~\ref{eq:CovPostDataBis} can be rewritten as
\begin{equation}
  \label{eq:CovPostData}
  \tilde{K}_{\mbx\mbx}^{-1} 
    = K_{\mbx\mbx}^{-1} + \FKtabT C_Y^{-1} \FKtab\, .
\end{equation}
Note that here, and in the rest of this paper, the notation $M_{\mbx\mbx}^{-1}$ denotes 
the inverse of the matrix $M_{\mbx\mbx}$ and not the $(\mbx,\mbx)$ block of the matrix 
$M^{-1}$. A more precise notation would be $(M_{\mbx\mbx})^{-1}$, not to be confused with 
$(M^{-1})_{\mbx\mbx}$. 

Eqs.~\eqref{eq:MeanPostData} and~\eqref{eq:CovPostData} were already obtained in
Ref.~\cite{DelDebbio:2021whr}, while Eq.~\eqref{eq:CovPostDataBis} provides an
alternative expression for the posterior covariance, and is the standard formulation in the context of Gaussian processes. 
Similarly, starting from the same prior and marginalizing with respect to 
$\mbf$ we can obtain the posterior for $\mbf^*$,
\begin{equation}
  \label{eq:PostDataGaussFs}
  p\left(\mbf^*| \theta,y\right)
  = \mathcal{N}\left(\mathbf{\tilde m}^*, \tilde{K}^*_{\mbx\mbx}\right)\, , 
\end{equation}
with
\begin{align}
  \label{eq:PostMeanStar}
  \tilde{\mbm}^* &= \mbm^* + K_{\mbx^*\mbx} \FKtabT 
    \covthdat^{-1} 
    \left(\mathbf{y} - \FKtab \mathbf{m}\right)\, ,\\
  \label{eq:PostCovStar}
  \tilde{K}_{\mbx^*\mbx^*} &= K_{\mbx^*\mbx^*} - 
    K_{\mbx^*\mbx} \FKtabT\, \covthdat^{-1} \, \FKtab K_{\mbx\mbx^*}\, .
\end{align}
Focusing on the corrections to the mean of the process due to Bayesian
inference,
\begin{align*}
  \label{eq:Delta1}
  \Delta\mathbf{m}   &= \mathbf{\tilde m} - \mathbf{m}\, , \\
  \Delta\mathbf{m^*} &= \mathbf{\tilde m^*} - \mathbf{m^*}\, ,
\end{align*}
we find
\begin{equation}
  \label{eq:MeanRelation}
  \Delta\mathbf{m^*} = K_{\mbx^*\mbx} K_{\mbx\mbx}^{-1} \Delta \mathbf{m}\, ,
\end{equation}
and 
\begin{equation}
  \label{eq:CovRelation}
  \tilde K_{\mbx^*\mbx^*} = K_{\mbx^*\mbx^*} -
    K_{\mbx^*\mbx} K_{\mbx\mbx}^{-1} K_{\mbx\mbx^*} + 
    K_{\mbx^*\mbx} K_{\mbx\mbx}^{-1}
    \tilde K_{\mbx\mbx} K_{\mbx\mbx}^{-1} K_{\mbx\mbx^*}\, .
\end{equation}

Let us emphasise once again that, in this approach, the values of the function
$f$ are stochastic variables, and the {\em information} that we can retrieve
about the function at the points $\mathbf{x^*}$ is precisely encoded in the
posterior probability distribution. Rather than finding {\em one} solution, we
find the probability distribution of the vector $\mathbf{f^*}$. This is
reminescent of what is done when bootstrapping a fit to the data: the posterior
distribution in this latter case is the distribution of fit results over the
bootstrap sample. 

\subsection{Inference for the Hyperparameters}
\label{sec:HPInference}
We now turn to the second term of Eq.~\eqref{eq:FullPosterior}, 
namely the posterior of the hyperparameters $\theta$ given the
data. Using Bayes' theorem we have
\begin{align}
  \label{eq:PosteriorHyper}
    p\left(\theta| y\right) =  
     \frac{p\left(y|\theta\right)p_{\theta}\left(\theta\right)}
      {\int d\theta\, p\left(y|\theta\right)p_{\theta}\left(\theta\right)}\,,
\end{align}
where $p_{\theta}\left(\theta\right)$ denotes the hyperparameters prior. The
likelihood $p\left(y|\theta\right)$ is proportional to the normalization of the
probability distribution  $p\left(\mbf|\theta,y\right)$ in Eq.~\eqref{eq:PosteriorF}, and as such can be computed 
integrating over $\mbf$.
Alternatively we can get its explicit expression by noticing that the
observed data are given by $y = \FKtab \, \mbf + \epsilon$ with
\begin{align}
    &\FKtab \, \mbf \sim \mathcal{N}\left(\FKtab\, \mbm, 
      \FKtab K_{\mbx \mbx} \FKtabT\right)\,, \\
    &\epsilon \sim \mathcal{N}\left(0, C_Y\right)\,,
\end{align}
and that therefore
\begin{align}
    y \sim \mathcal{N}\left(\FKtab\, \mbm, 
    \covthdat\right)\,,
\end{align}
with $\covthdat$ defined in Eq.~\eqref{eq:covthdatDef}.
In both cases one finds
\begin{align}
  \label{eq:data_given_theta}
    p\left(y | \theta\right) = 
     \frac{e^{-\frac{1}{2}\,\left(y-\FKtab \mbm \right)^T\covthdat^{-1}\left(y-\FKtab \mbm\right)}}
     {\sqrt{\det \left[2\pi \covthdat\right]}}\,.
\end{align}

Note that the inference for the model, which yields the posteriors $p\left(\mbf | \theta, y\right)$ and
$p\left(\mbf^* | \theta, y\right)$, is completely analytical. 
This second inference step, which determines the posterior $p\left(\theta| y\right)$, 
in general cannot be solved
analytically: because of the hyperparameters appearing in the square root in 
Eq.~\eqref{eq:data_given_theta},
the denominator of Eq.~\eqref{eq:PosteriorHyper} cannot be computed analytically any longer.
Moreover $p(y|\theta)$, as a function of $\theta$, 
in general is not a conveniently analyzable probability density function 
that we know how to sample from.
Therefore, this step has to be addressed as a standard inference problem. We can
then follow two approaches:
\begin{itemize}
    \item select the hyperparameter values as the mode of the posterior $p\left(\theta| y\right)$:
    \begin{align}
        \theta_{\text{MAP}} = \text{arg max}_{\theta} \,p\left(\theta| y\right),
    \end{align}
    \item use an MCMC algorithm to sample from the posterior $p\left(\theta| y\right)$. 
\end{itemize}
Using this second approach the uncertainty due to hyperparameter selection is
incorporated into the final PDF uncertainty, and PDFs fitting is reduced to a Monte
Carlo problem.

While finding $\theta_\text{MAP}$ is computationally less demanding than MCMC, we opt
for the full inference because of our focus on uncertainty quantification.
Stopping at a single ``best'' value can dramatically alter the posterior variance, see,
e.g., \cite[Fig.~18.18, p.~713]{murphy2023}.

\subsection{The quadratic case}
\label{sec:QuadCase}
So far we have considered the case in which the theoretical predictions
are linear in $f_i$, according to Eq.~\eqref{eq:LinThGrid}.  
In this section we discuss what changes when we consider observables
with quadratic dependence on $\mbf$. 
%
Denoting the FK-table for a quadratic observable $T^{\text{quad}}$ as $\widehat{\FKtab}$,
Eq.~\eqref{eq:LinThGrid} becomes
\begin{equation}
  \label{eq:QuadThGrid}
  T^{\text{quad}}_I = \sum_{i,j=1}^{N} \widehat{\FKtab}_{Iij} f_i f_j\,.
\end{equation}
The prior given in Eq.~\eqref{eq:PriorEpsilonDistr} remains unchanged, but conditioning
now results in
\begin{align}
  \mbf^T \widehat{\FKtab} \mbf + \epsilon = y\,, 
\end{align}
so that, following the same steps as in Sec.~\ref{sec:GPModel},
the posterior $p(\mbf|\theta,y)$ is now given by
\begin{align}
  p(\mbf|\theta,y) \propto \exp\left\{-\widehat{S}(\mbf; \theta, y)\right\}\,,
\end{align}
with
\begin{align}
  \label{eq:QuadLogLoss}
  \widehat{S}(\mbf; \theta, y) = 
    \frac12 \left\{
      (\mathbf{f}-\mathbf{m})^T
      K_{\mbx\mbx}^{-1} 
          (\mathbf{f}-\mathbf{m}) +
          (\mbf^T \widehat{\FKtab} \mbf - y)^T 
          C_Y^{-1} (\mbf^T \widehat{\FKtab} \mbf - y)
    \right\}\,.
\end{align}
The full posterior $p\left(\mbf,\theta |  y\right)$ can be written using Bayes' theorem as
\begin{align}
  \label{eq:FullPosteriorQuad}
  p\left(\mbf,\theta |  y\right) = 
  \frac{\exp\left\{-\widehat{S}(\mbf; \theta, y)\right\}
   p_{\theta}\left(\theta\right)}
   {\int d\theta d\mbf  \exp\left\{-\widehat{S}(\mbf; \theta, y)\right\}
   p_{\theta}\left(\theta\right) }\,.
\end{align}
By multiplying numerator and denominator by the marginal likelihood
\begin{align}
\label{eq:marglikelihood}
  p\left(y|\theta\right) \propto \int d\mbf  \exp\left\{-\widehat{S}(\mbf; \theta, y)\right\}
\end{align}
we can recast Eq.~\eqref{eq:FullPosteriorQuad} in the same form as Eq.~\eqref{eq:FullPosterior}
with
\begin{align}
  & p\left(\mbf| \theta,y\right) = \frac{\exp\left\{-\widehat{S}(\mbf; \theta, y)\right\}}
  {p\left(y|\theta\right)} \quad \text{and} \quad
  p\left(\theta | y \right) = 
  \frac{p\left(y|\theta\right) p_{\theta}\left(\theta\right)}
   {\int d\theta p\left(y|\theta\right)
   p_{\theta}\left(\theta\right)}\,.
\end{align}
The difference with respect to the linear case is that Eq.~\eqref{eq:QuadLogLoss} 
is not quadratic in $\mbf$. It follows that the posterior $p(\mbf|\theta,y)$ is not a Gaussian any longer
and the likelihood $p(y|\theta)$ cannot be computed analytically. 

Both the inference on the parameters $\mbf$ and on the hyperparameters $\theta$ 
has therefore to be performed at the same time by running a MCMC algorithm 
starting from Eq.~\eqref{eq:FullPosteriorQuad}. 
No simplifications occur, and the dimension of the Monte Carlo 
to run, corresponding to $\text{dim} \left(\theta\right)$ in the linear case, is now
$\text{dim} \left(\mbf\right) + \text{dim} \left(\theta\right)$.
This very same approach would work for a generic functional of $f$.

%% file: example_ns.tex
\section{Gaussian Processes for PDFs}
\label{sec:examples}
In the following we apply the formalism described in the previous section 
to two concrete examples. We consider only the simpler case of observables linear in $\mbf$,
where the inference on the parameters can be done analytically. 
We first show how, by a suitable choice of the prior, we can implement 
known physical constraints such as the kinematic limit, sum rules and small-$x$ behaviour.
We then give a complete example of the workflow, by determining the nonsinglet triplet PDF 
\[ 
    T_3 = u^+ - d^+\,,
\]
using a set of synthetic data, first for DIS structure functions, and then for lattice 
equal-time correlators.

\subsection{Prior distribution for $T_3$}
\label{sec:priorT3}
When fitting PDFs we can work in the so-called evolution basis
with six non-singlet quark distributions, 
\begin{equation}
    \label{eq:TsAndVs}
    T_{\sfa}(x)\, , \, \sfa=3,8,15\, , \quad
    V_{\sfa}(x)\, , \, \sfa=3,8,15 \, , 
\end{equation}
the quark singlet distribution, $\Sigma(x)$, and the gluon distribution, $g(x)$.
In the following we will be interested in the  
nonsinglet triplet distribution $T_3$, to which we associate a GP with zero
mean and kernel $k$ 
\begin{equation}
    \label{eq:TPriors}
    T_{3} \sim \GP(0,k)\,.
\end{equation}

The choice of the GPs that define the prior distributions needs to reflect the
knowledge of any physical property of the system.
We use here a Gibbs kernel~\cite[p.~93]{books/lib/RasmussenW06},
\begin{align}
    \label{eq:gibbs_kernel}
    k\left(x,y\right) = \sigma^2 \sqrt{\frac{2 l\left(x\right)l\left(y\right)}{l^2\left(x\right) 
    + l^2\left(y\right) }} 
    \exp\left[-\frac{\left(x-y\right)^2}{l^2\left(x\right) + l^2\left(y\right)}\right]\,
\end{align}
with
\begin{align}
    l\left(x\right) = l_0 \times \left(x + \delta\right)\,.
\end{align}
The quantities $\sigma$ and $l_0$ are hyperparameters of the GP, while
$\delta$ is a small fixed number which regularizes $k$ when $x,y \rightarrow
0$. 
This choice ensures that when approaching the small-$x$ domain the correlation length 
decreases linearly in $x$, reflecting the little knowledge we have in this kinematic region.
Note that 
\begin{align}
    k\left(x,x\right) = \sigma^2\,,
\end{align}
which implies a constant amplitude of the kernel on the full $x$ domain.
Since we do know something regarding the power behaviour of the PDF at small-$x$,
it would be convenient to encode it in the prior.
This can be done by introducing an additional hyperparameter $\alpha$ and by rescaling the kernel 
\begin{align}
    \label{eq:smallx_behaviour}
    k\left(x,y\right) \mapsto \phi\left(x\right)k\left(x,y\right)\phi\left(y\right)\,,
\end{align}
with
\begin{align}
    \label{eq:rescaling_function}
    \phi\left(x\right) = x^{\alpha}\,,
\end{align}
so that for the rescaled kernel
\begin{align}
    k\left(x,x\right) \propto x^{2\alpha}\,.
\end{align}
In the case of $T_3$, the PDF has to be integrable in $x=0$, which can be imposed by choosing
$\alpha \in \left(-1,0\right]$.
More properties can be implemented in the prior, such as sum rules and kinematic limit,
discussed in Appendix~\ref{app:prior_PDF},~\ref{app:kinlim}.

It should be kept in mind that the choice of the kernel is crucial, 
and that, when limited experimental data are available, different kernels 
lead to different results. For the sake of this paper, which aims at presenting the 
main ideas of the methodology in simple terms, we limit our study to the case of the Gibbs kernel.
However, when moving to more complex studies which aim to be used for phenomenology, 
different choices should be explored, by testing different kernels or defining new kernels suitable 
for the specific problem of PDF determination.
Here follow two ways a particular choice of kernel could turn out wrong in our case. 
First, the kernel almost completely determines the extrapolation behavior: 
in this case a mistaken assumption cannot be corrected by the data. 
Second, and more subtly, it is easy to inadvertently define an ill-conditioned kernel; in other words, 
a kernel which for practical matters behaves as a finitely parametric model, or that is still ``fat'' in infinite dimensions, but induces non-zero probability only on some overly specific functions. 
The textbook example of bad kernel is the exponential quadratic $e^{-(x-y)^2}$, 
widely used for its simplicity, yet encoding a strong prior; 
the Gibbs kernel is a variant of it, and so we expect it to have similar problems. 
Sure that (our own) potential future works will coast along with the Gibbs kernel by inertia, we say, 
Reader: heed our bootless warning. 

\subsection{Example 1: $T_3$ from BCDMS data}

Considering DIS on a proton target, the NNLO theory prediction
for the structure function $F_2^p$ is
\begin{align}
    F^p_2 = C_g \otimes g + C_{\Sigma} \otimes \Sigma 
        + C_{T_3} \otimes T_3 + C_{T_8} \otimes T_8 + C_{T_{15}} \otimes T_{15}\,,
\end{align}
where $g,\,\Sigma,\,T_3,\,T_8,\, T_{15}$ denote PDF flavors in the so-called 
evolution basis and $C_i$ the corresponding Wilson coefficients. 
Considering a neutron target instead and assuming isoscalarity, 
the neutron PDFs are just the same, except for $u, \bar{u}$ and $d, \bar{d}$, which are 
exchanged. 
Since $T_3 = u^+ - d^+$ it follows that the neutron structure function $F_2^n$ can be written as
\begin{align}
    F^n_2 = C_g \otimes g + C_{\Sigma} \otimes \Sigma 
        - C_{T_3} \otimes T_3 + C_{T_8} \otimes T_8 + C_{T_{15}} \otimes T_{15}\, , 
\end{align}
with the same Wilson coefficients as in the proton case. Considering a deuterium
target, for a generic flavor the corresponding nuclear PDF is
\begin{align}
    f^d_i = \frac{1}{2}\left(f^p_i + f^n_i\right)\,.
\end{align}
The deuterium structure function is therefore given by averaging
the ones for proton and neutron, getting 
\begin{align}
    F^d_2 = C_g \otimes g + C_{\Sigma} \otimes \Sigma + C_{T_8} \otimes T_8 + C_{T_{15}} \otimes T_{15}\,. 
\end{align}
Hence we can define the observable 
\begin{align}
    \label{eq:NS_observable}
    F^p_2 - F^d_2 = C_{T_3} \otimes T_3\, ,
\end{align}
which is expressed as the convolution of the  Wilson coefficient $C_{T_3}$ with
just one PDF, \viz\ $T_3$. The determination of $T_3$ using $F^p_2 - F^d_2$ only
involves one flavor and one FK table, making it an ideal testbed for the 
method. 

\paragraph{Data and FK table}
Rather than considering real experimental data, we will consider pseudo-data constructed 
from a known underlying law. This will allow us to test how well 
the methodology is able to reconstruct the input model (see discussion in Sec.~\ref{sec:discussion}).
Starting from the datasets BCDMSP and BCDMSD presented in Ref.~\cite{Benvenuti:1989rh},
pseudo-data are generated by identifying points for 
$F^p_2$ and $F^d_2$ having the same values of
the kinematic variables, and taking their difference, which yields a total of
333 points. Following the standard procedure in PDF fits based on factorization, we apply 
kinematic cuts excluding datapoints where power suppressed corrections could be relevant, leaving 
248 points in our analysis. 
For the experimental error $C_Y$, 
we consider the full experimental covariance for the observable $F^p_2-F^d_2$, 
\begin{align}
    C_Y = \cov\left[F^p_2, F^p_2\right] + \cov\left[F^d_2, F^d_2\right]
    - 2 \cov\left[F^p_2, F^d_2\right]\, ,
\end{align}
which is computed using the publicly-available experimental information.
As underlying law we could use any functional form we like. 
To consider a realistic scenario we take the central value of the recent PDF release NNPDF4.0.
We denote as $y_0$ 
the data generated from the underlying law $\mbf_0$ using the corresponding $\FKtab$ table
\begin{align}
\label{eq:ytrue}
    y_0 = \FKtab \, \mbf_0\,.
\end{align}
The corresponding experimental measurements $y$ entering 
Eqs.~\eqref{eq:DataWithErr},~\eqref{eq:data_given_theta} are built as 
\begin{align}
    \label{eq:yobs}
    y = y_0 + \eta\,, \quad \text{with} \quad \eta ~ \sim \mathcal{N}\left(0,C_Y\right)\,.
\end{align}

\begin{figure}[h!]
    \center
    \includegraphics[scale=0.4]{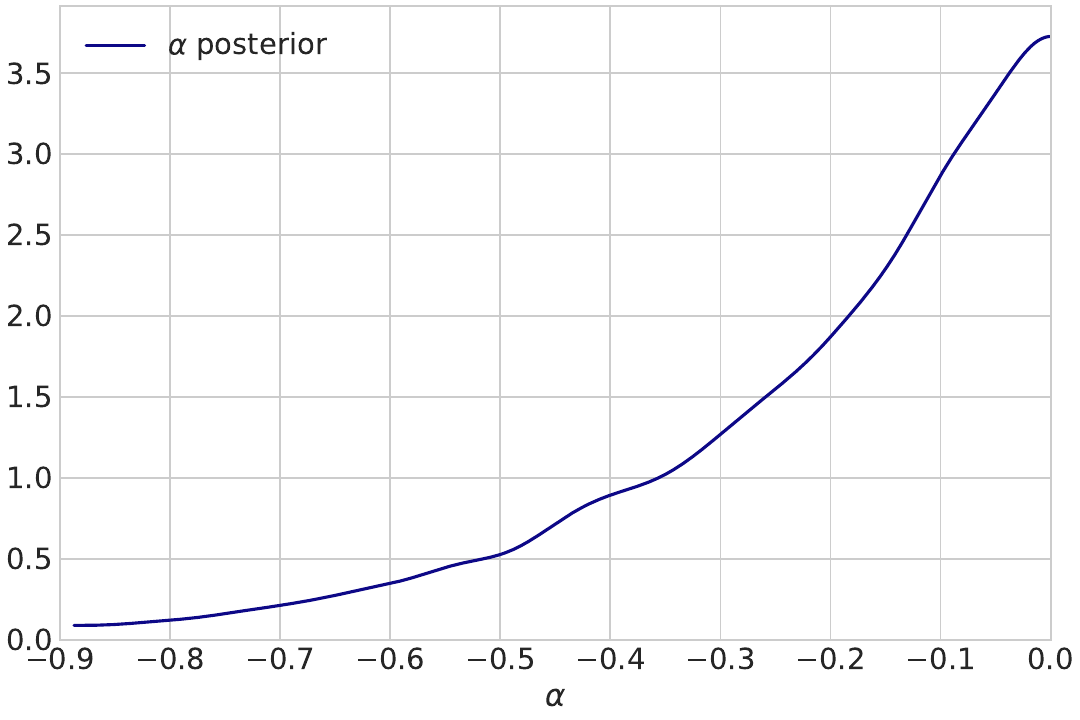} 
    \includegraphics[scale=0.4]{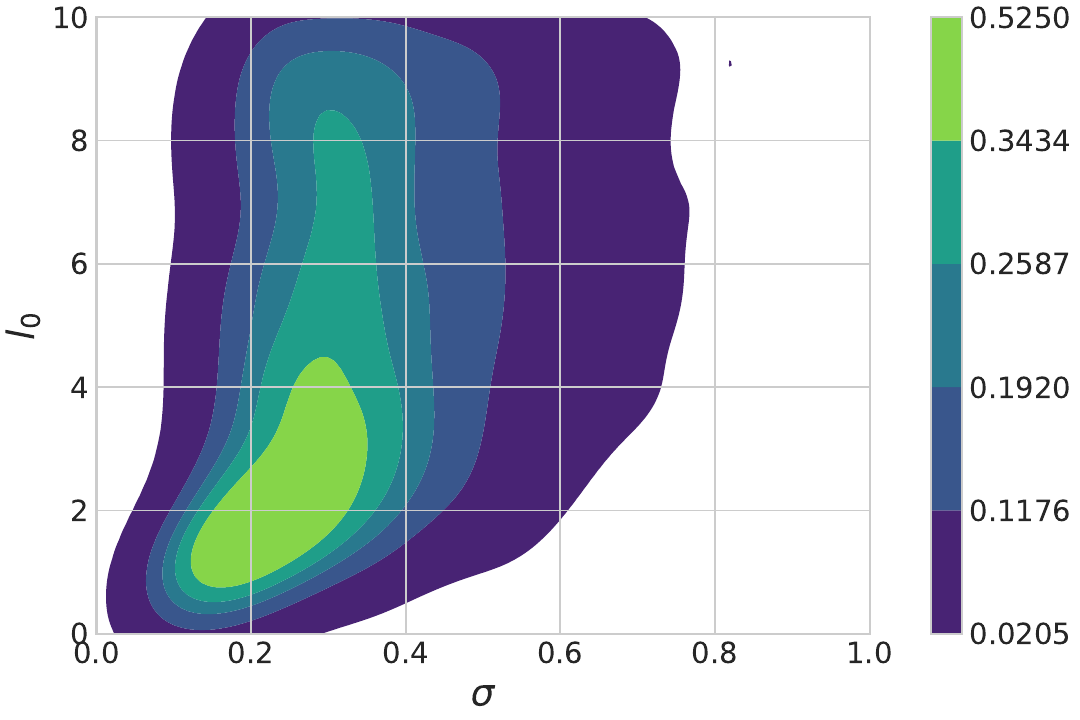}
    \includegraphics[scale=0.4]{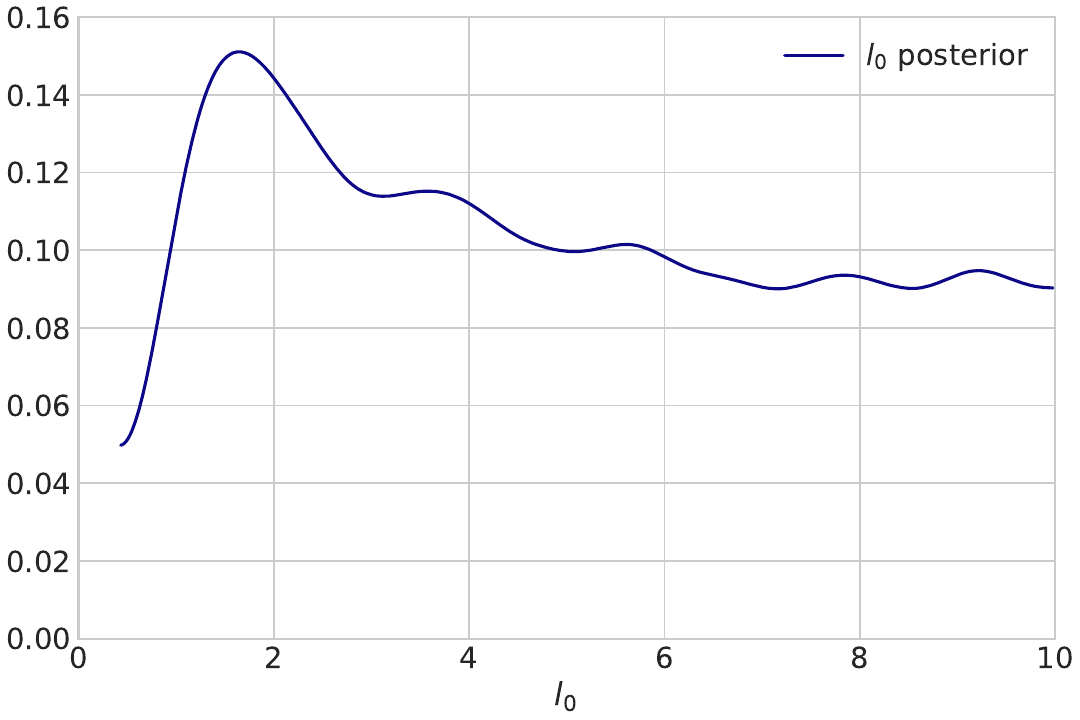} 
    \includegraphics[scale=0.4]{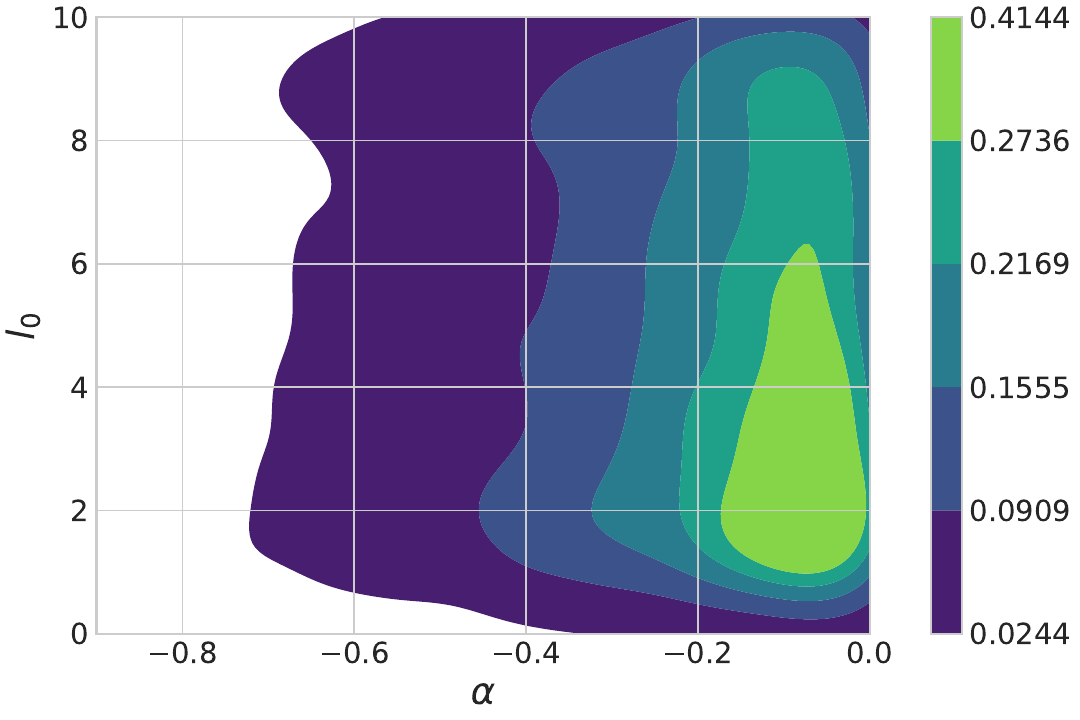}
    \includegraphics[scale=0.4]{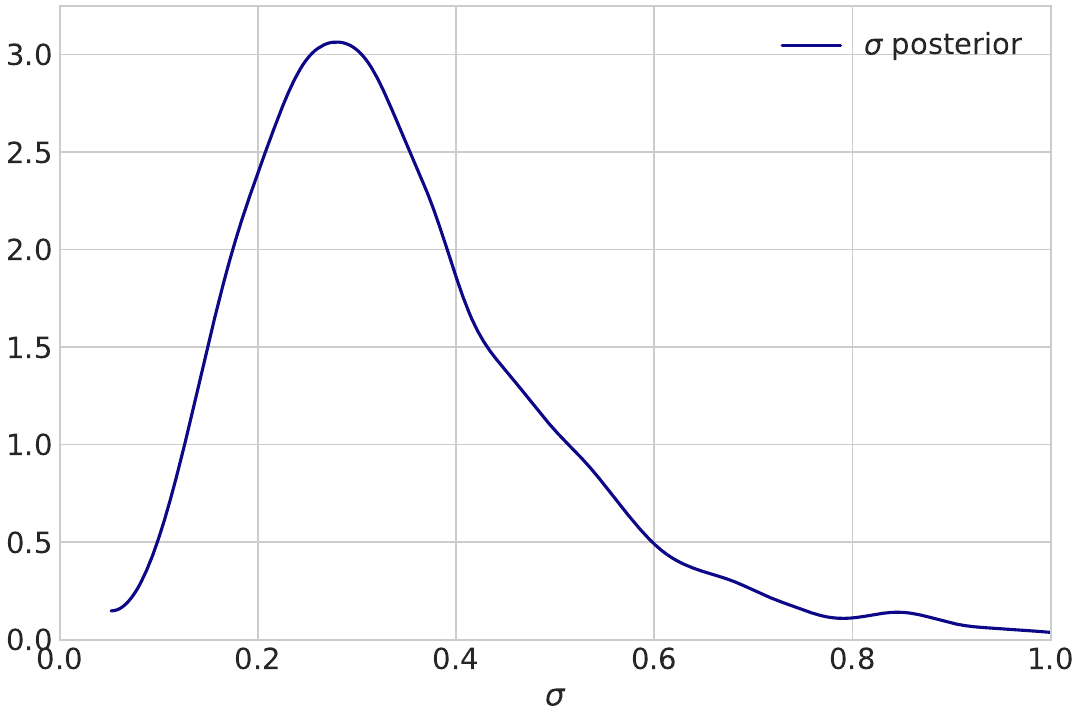} 
    \includegraphics[scale=0.4]{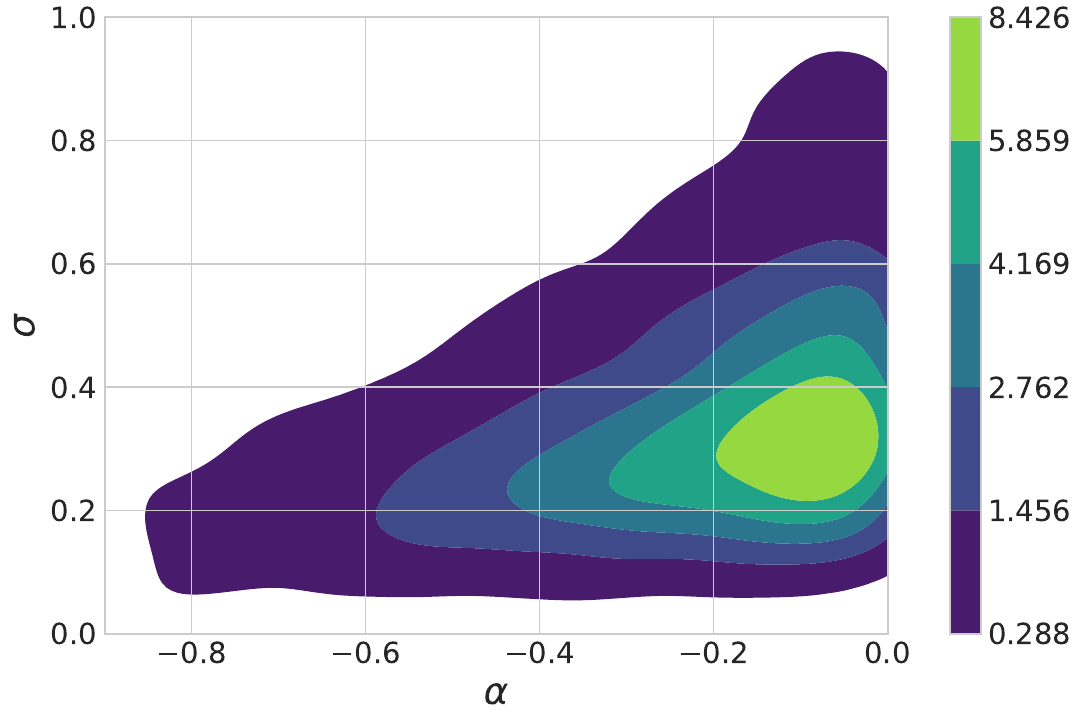}
    \caption{\small 1-dimensional (left panel) and 2-dimensional (right panel) 
    posteriors of the hyperparameters $\alpha$, $l_0$ and $\sigma$.
    The hyperparameter $\sigma$ is characterized by a sharply peaked posterior
    located around $\sigma\sim 0.25$ which quickly decays to zero 
    (for this reason in the posterior plot only the region $\left(0,1\right)$ is shown,
    even if the support of the prior is $\left(0,10\right)$); 
    $\alpha$ tends to sit closer to $0$, with a slow decay for smaller values towards $-1$;
    the $l_0$ posterior discards the smaller values of the correlation length, 
    it shows a peak for $l_0\sim 1.7$ and then remains fairly constant.  
    \label{fig:hp_posteriors}
    }
\end{figure}

\paragraph{Hyperparameters inference}
Hyperparameters inference is the first step of the procedure and is carried on as described in 
Sec.~\ref{sec:HPInference}.
The hyperparameters entering the analysis are $\alpha$, $l_0$ and $\sigma$.
As a prior we choose a flat\footnote{Flat priors are not a good default in general. 
Here, with only three free hyperparameters, the choice of prior does not matter much; 
but with the additional model complexity we expect to employ for the full PDF analysis, 
it will be worth making a reasoned choice.} distribution
having support $\left(0,10\right)$ in the case of $l_0$ and $\sigma$,
and $\left(-1,0\right)$ in the case of $\alpha$, 
in order to ensure the integrability of $T_3$ at small-$x$.
To produce results we have run an MCMC algorithm using 
the Python package {\tt PyMC} \cite{pymc2023}: the NUTS sampler is run on 4 
independent chains, which are merged after thermalization in a unique set of samples.
The posteriors for the three hyperparameters are plotted in Fig.~\ref{fig:hp_posteriors}.
Starting from flat priors, the inference based on the available data generates 
non-trivial posterior distributions. We will comment more extensively on these results later. 

\begin{figure}[h!]
    \center
    \includegraphics[scale=0.4]{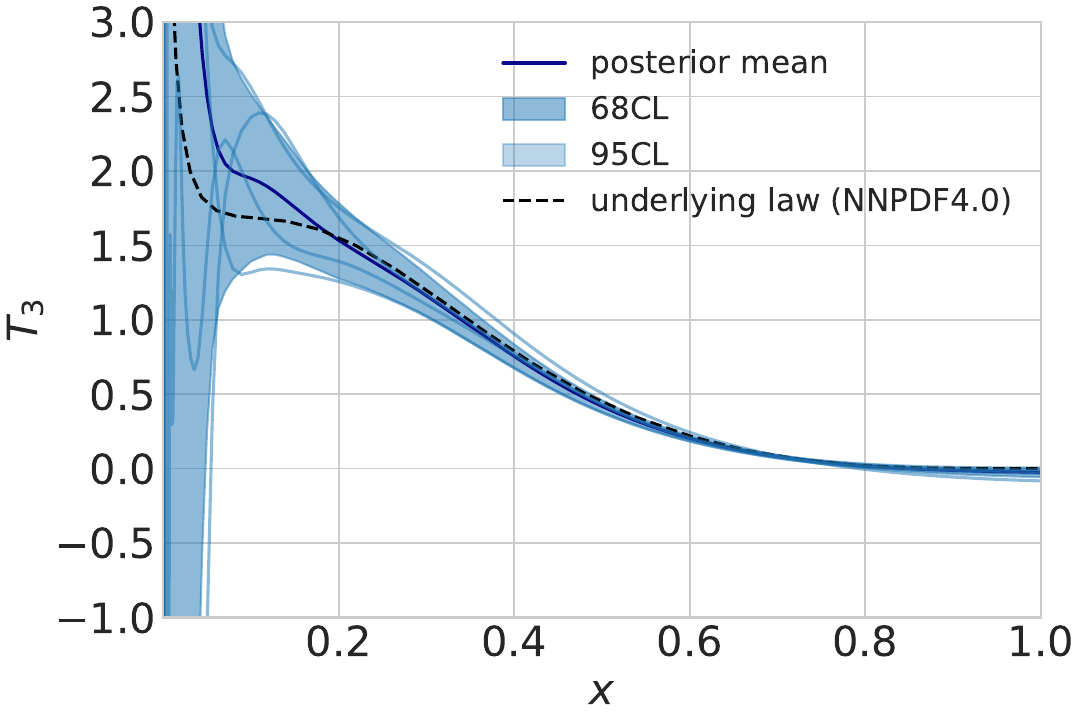}
    \includegraphics[scale=0.4]{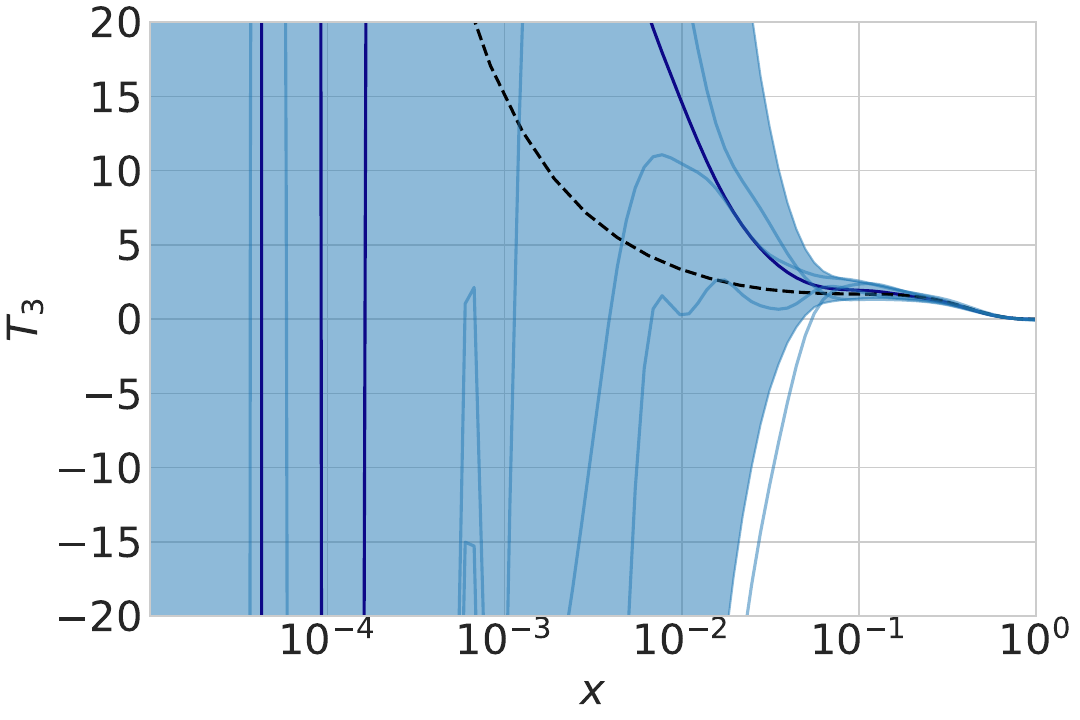}
    \caption{\small Samples from  $p\left(\mbf^*,\alpha,l_0,\sigma | y,C_Y\right)$
    plotted in linear (left panel) and log (right panel) scale.
    The dark blue line represent the mean of the distribution, while the black dotted line the 
    input PDF NNPDF4.0 used to generate pseudo-data. 
    The shaded regions represent the 68CL and 95CL intervals, and in light blue we plot a few representative samples
    from the distribution. 
    The posterior displays a smaller variance in the regions sensitive to experimental data, 
    and an increasing spread in the small and large-$x$ extrapolation regions, 
    where the results are mostly determined by the chosen prior. 
    \label{fig:mcT3}
    }
\end{figure}

\paragraph{Gaussian inference and results}
Having determined the posterior of the hyperparameters $p\left(\theta|y,C_Y\right)$, we can 
generate an ensemble of hyperparameters by sampling this distribution. 
For each hyperparameters sample, the posterior of the parameters $p\left(\mbf^*|\theta,y,C_Y\right)$
can be computed analytically, and a Gaussian replica can be drawn from it. This two-step procedure
produces exactly a sample from $p\left(\mbf^*,\theta| y, C_Y\right)$.
In Fig.~\ref{fig:mcT3} we show the final results, obtained by sampling
from $p\left(\mbf^*,\theta| y, C_Y\right)$.

\subsection{Example 2: $T_3$ from lattice data}

\begin{figure}[h!]
    \center
    \includegraphics[scale=0.4]{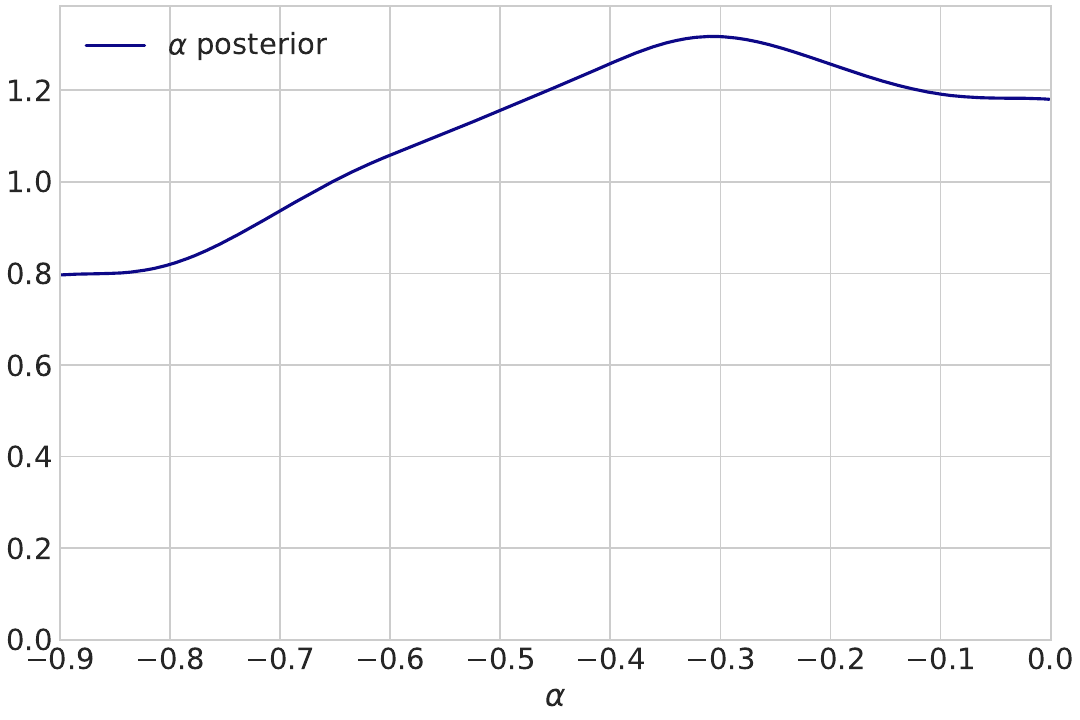} 
    \includegraphics[scale=0.4]{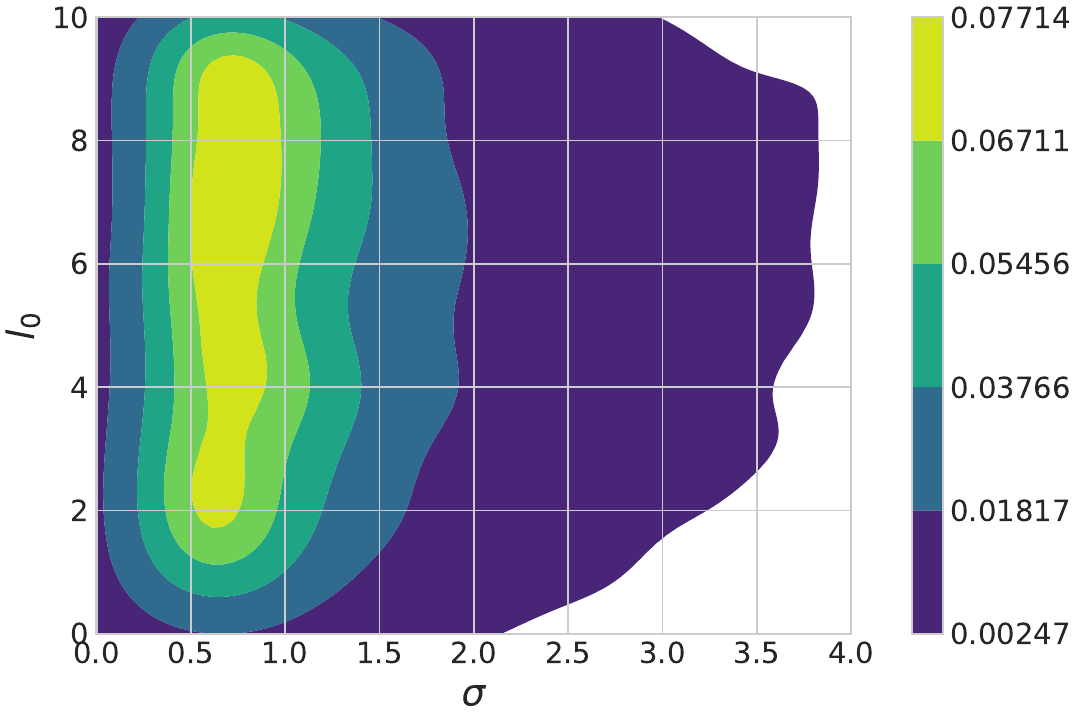}
    \includegraphics[scale=0.4]{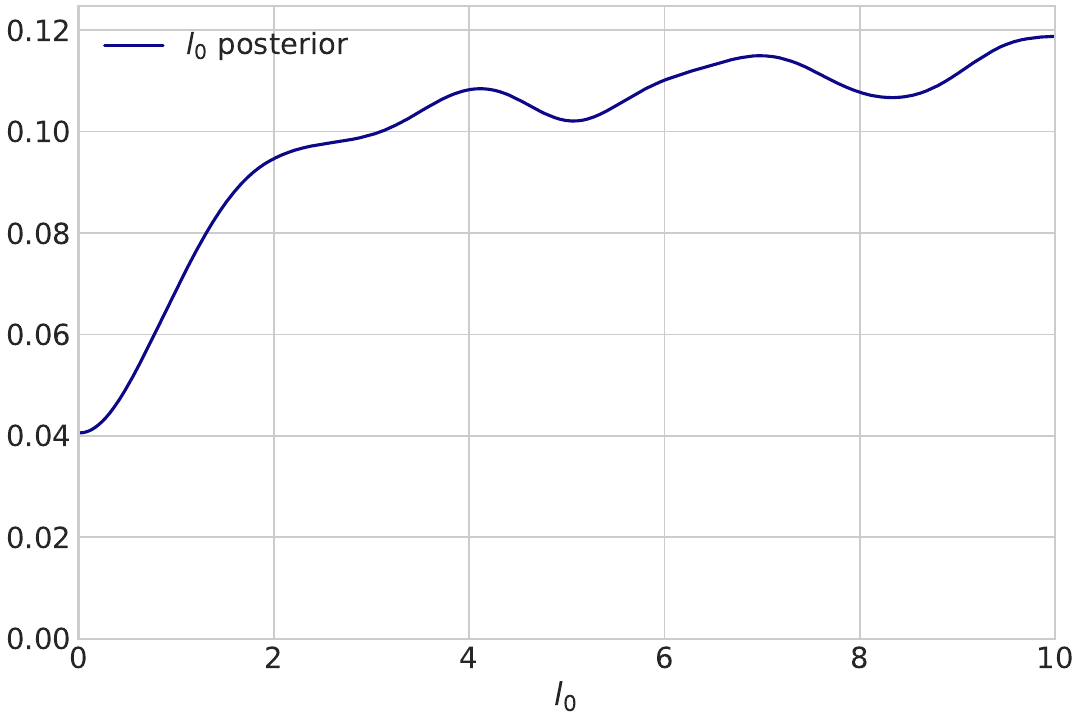} 
    \includegraphics[scale=0.4]{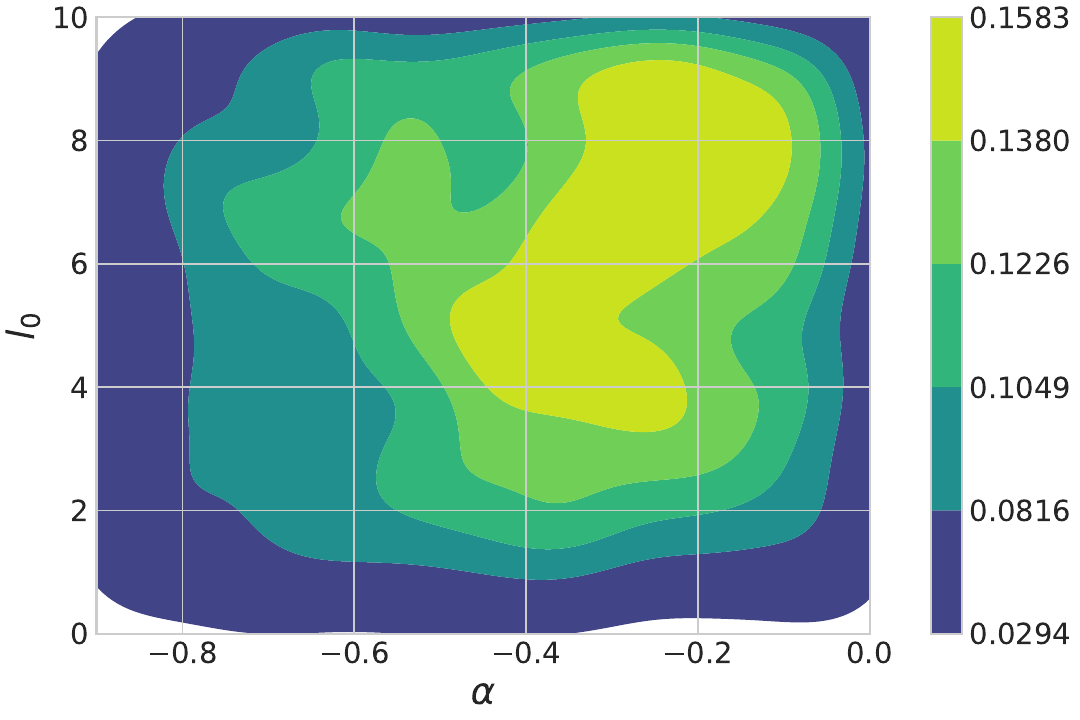}
    \includegraphics[scale=0.4]{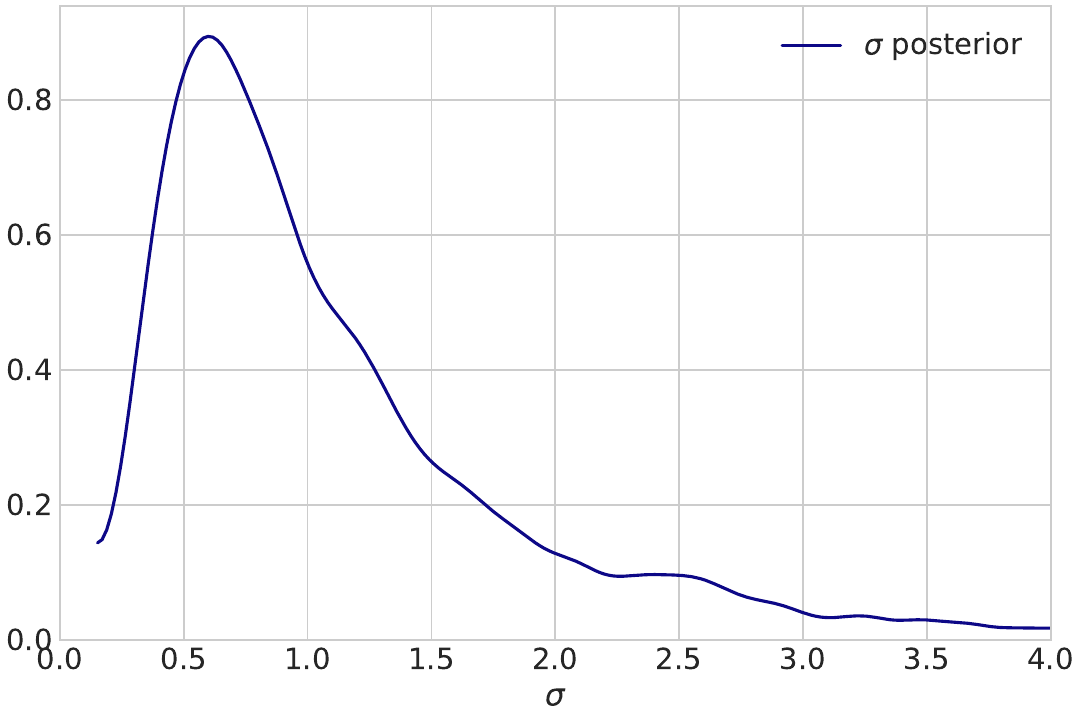} 
    \includegraphics[scale=0.4]{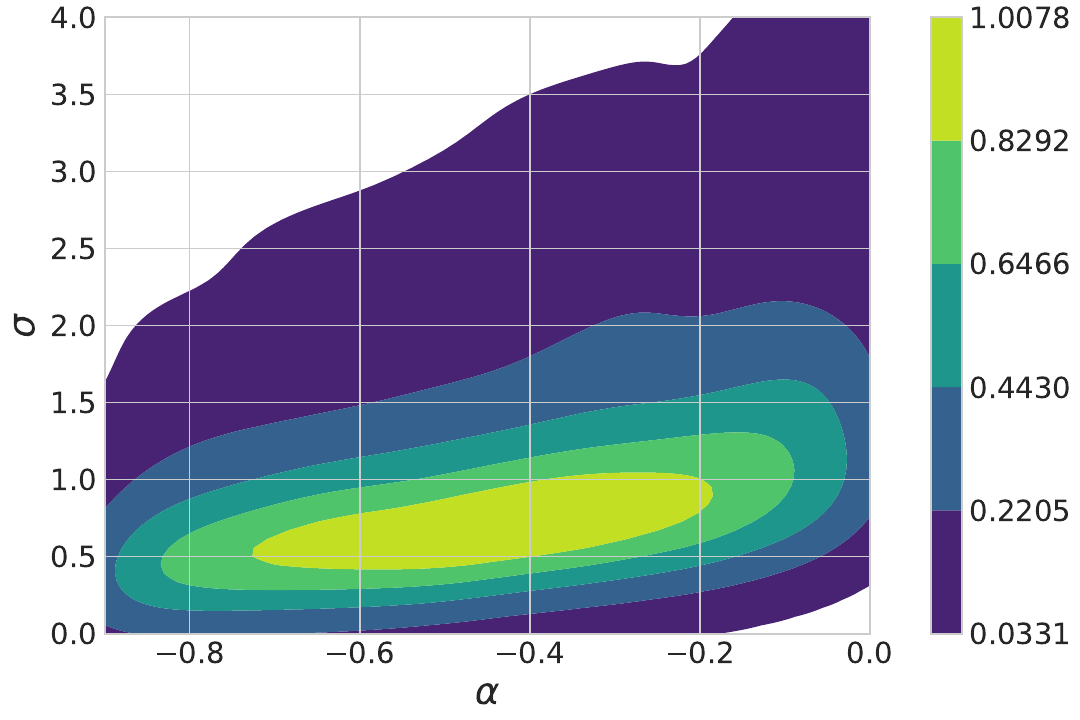}
    \caption{\small Same as Fig.~\ref{fig:hp_posteriors} for the case of lattice pseudo-data.
    While the hyperparameter $\sigma$ is characterized by a sharply peaked posterior distribution,
    both $\alpha$ and $l_0$ show fairly flat posteriors, with the one for $l_0$ only penalizing 
    smaller values of the correlation length.
    \label{fig:hp_posteriors_lat}
    }
\end{figure}

\begin{figure}[h!]
    \center
    \includegraphics[scale=0.4]{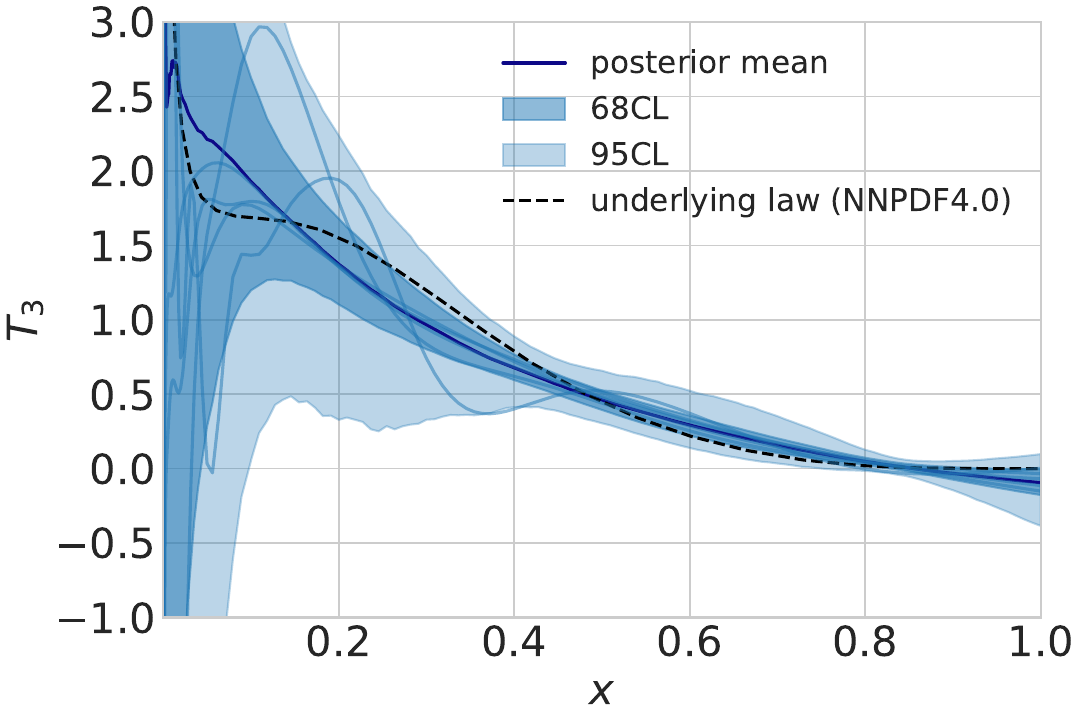}
    \includegraphics[scale=0.4]{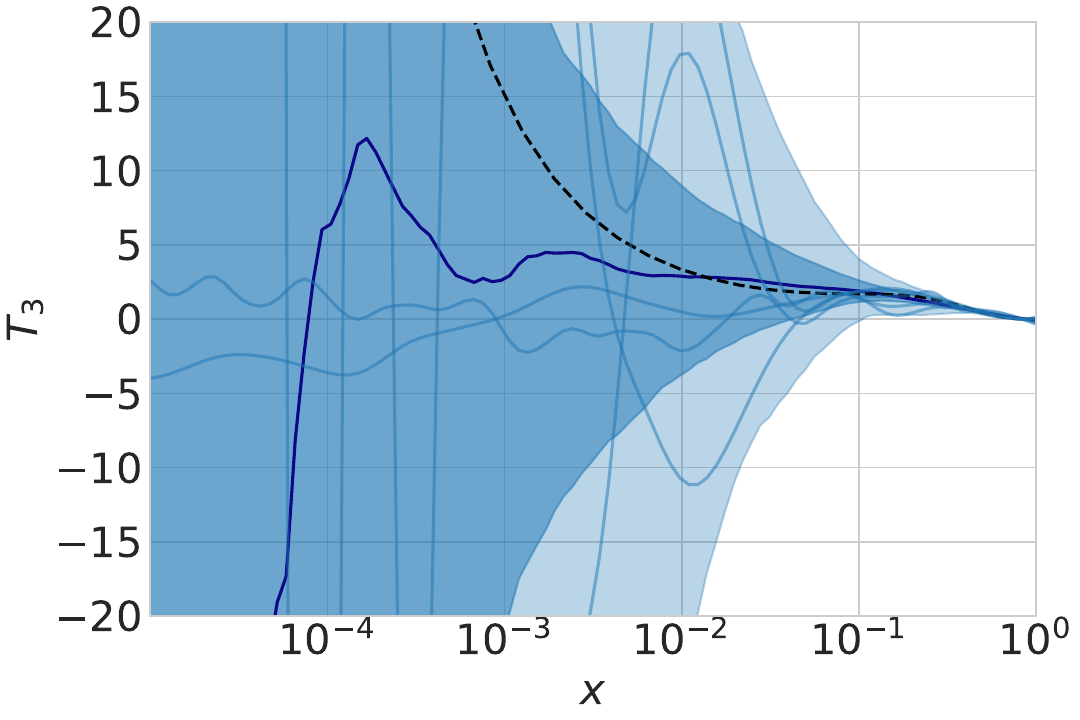}
    \caption{\small Same as Fig.~\ref{fig:mcT3} for the case of lattice pseudo-data.
    As in the case of BCDMS data, smaller error are observed in the kinematic regions
    which are sensitive to the available data. 
    In general, by comparing this plot with Fig.~\ref{fig:mcT3}, it is clear how lattice data
    provide less stringent constraints on $T_3$ than BCDMS. Nevertheless the input PDF
    is still well reconstructed by the posterior.
    \label{fig:mcT3_lat}
    }
\end{figure}

The same kind of inverse problem as the one presented in the previous section
is found when reconstructing PDFs from a discrete set of values 
for lattice equal-time correlators~\cite{Cichy:2019ebf,DelDebbio:2020rgv,Karpie:2019eiq}.
Following Ref.~\cite{DelDebbio:2020rgv}, we can reconstruct the distribution $T_3$ from a set of data for 
reduced Ioffe-time pseudodistributions~\cite{Radyushkin:2017cyf}.
Denoting the latter as $\mathcal{M}\left(\nu, z_3^2\right)$, its imaginary part is related to $T_3$ 
by the integral relation 
\begin{align}
    \label{eq:lattice_obs}
    \text{Im}\left[\mathcal{M}\right] = \int_0^1 dx\,C^{\text{Im}}\left(x\nu, \mu^2z_3^2\right)
    T_3\left(x,\mu^2\right)\,.
\end{align}
Also in this case we consider pseudo-data: central values are built according to Eq.~\eqref{eq:lattice_obs}
using the analytical expression and the kinematic values 
described in Ref.~\cite{DelDebbio:2020rgv} and NNPDF4.0 as input PDF set.
This gives a total of 48 points in the $\left(\nu,z_3^2\right)$ plane. 
The covariance matrix $C_Y$ in this case is given by the uncertainties coming 
from the actual lattice simulation and we use here
the covariance described in Ref.~\cite{DelDebbio:2020rgv}.
Also in this case, Eq.~\eqref{eq:lattice_obs} is implemented as per Eq.~\eqref{eq:LinThGrid}
by means of suitable FK tables.
We repeat the same steps as in the previous section, 
starting from the same prior for $T_3$ and changing only the FK tables and data entering 
the framework. 
The posterior for the hyperparameters and the resulting PDF are plotted in
Figs.~\ref{fig:hp_posteriors_lat},~\ref{fig:mcT3_lat}.


\subsection{Discussion}
\label{sec:discussion_res}
Looking at the posteriors plotted in Figs.~\ref{fig:hp_posteriors}, and \ref{fig:hp_posteriors_lat},
we notice how, even though we started from flat priors, the inference based on the available data
has generated non-trivial posterior distributions, with more or less sharp peaks depending 
on the specific case we consider: while BCDMS data give fairly peaked distributions, lattice data seem to 
allow a broader range of hyperparameter values, expecially in the case of $\alpha$ and $l_0$. 
As pointed out in Sec.~\ref{sec:HPInference}, the corresponding uncertainty is included in the final 
results: by sampling from the full posterior 
$p\left(\mbf^*,\theta |  y,C_Y\right)$ the PDF error plotted in Figs.~\ref{fig:mcT3}, and \ref{fig:mcT3_lat},
includes the component due to different possible values of the hyperparameters. 
Overall, the posterior distributions of the hyperparameters
plus the Gaussian sampling performed according to the posterior covariance matrix 
give a broader distribution at the level of the final PDF in the case of the lattice data.
Inspecting Fig.~\ref{fig:mcT3} a number of qualitative considerations can be done:
in the kinematic region sensitive to the data the input PDF $\mbf_0$ is reconstructed
with a small error, while in the small- and large-$x$ extrapolation region, where no experimental information
is available, the posterior strongly depends on the prior we chose in the first place.
As discussed in Sec.~\ref{sec:priorT3} we incorporate in the prior a behavior giving 
larger error which yet is compatible with integrability properties of $T_3$. 
Lacking of any small-$x$ experimental information, 
this is what we find back in the posterior at small values of $x$ (Fig.~\ref{fig:mcT3}, right panel).  
Similarly for $x > 0.8$ we enter the large-$x$ extrapolation region, which is reflected by an increase of the 
error band visible in the left panel of Fig.~\ref{fig:mcT3} 
(see Appendix~\ref{app:kinlim}).
Similar qualitative considerations can be done for the lattice data case,
by looking at the plot in Fig.~\ref{fig:mcT3_lat}.  

In the next section we will make the discussion around these results more quantitative, 
showing how different components entering the final error can be identified and by introducing
different metrics to validate the methodology.

%% file: closure.tex
\section{Uncertainties and validation}
\label{sec:discussion}

In the following we discuss the final uncertainty of the result,
and identify different components associated to the experimental and reconstruction error, 
the latter being associated to the ill-posed nature of inverse problems.
We then discuss the extent to which the underlying model used to generate pseudo-data is reconstructed.

\subsection{Decomposition of the posterior covariance matrix}

\paragraph{Vanishing experimental error}
Let us first consider the scenario of no experimental error, \ie\ let us 
assume that the experimental measurements reproduce the true data with no error.
From Eqs.~\eqref{eq:ytrue},~\eqref{eq:yobs} it follows
\[
y = y_0 = \FKtab \mbf_0 \,.
\]
Following Sec. 3.1.1 of Ref.~\cite{10.1093/gji/ggz520}, 
let us define the resolution kernel as
\begin{align}
    R^{(0)}_{\mbx^*\mbx} = K_{\mbx^*\mbx}\, (\fk)^T \left[(\fk) K_{\mbx\mbx} (\fk)^T\right]^{-1}
    (\fk)\, .
\end{align}
Eq.~\eqref{eq:PostMeanStar} can be rewritten as
\begin{align}
    \label{eq:mean_posterior_T3_1}
    \tilde{\mbm}^* - \mbm^*  
        &= R^{(0)}_{\mbx^*\mbx} \left(\mbf_0 - \mbm\right)
\end{align}
which, for $\mbm^* = \mbm = 0$, reduces to 
\begin{align}
    \label{eq:mean_posterior_T3_2}
    \tilde{\mbm}^* = R^{(0)}_{\mbx^*\mbx}\, \mbf_0\,.
\end{align}
Eq.~\eqref{eq:mean_posterior_T3_2} shows that the result of Bayesian inference
is a smeared version of the ``true" answer $\mbf_0$, with the smearing kernel
given by $R^{(0)}_{\mbx^*\mbx}$. The difference between the mean value of
the posterior and the underlying law is
\begin{align}
    \label{eq:bias_function_space}
    \tilde{\mbm}^* - \mbf^*_0 = R^{(0)}_{\mbx^*\mbx} \mbf_0 - \mbf^*_0\, .
\end{align}
We can further specialize the
discussion by considering the case $\mbx^* = \mbx$ (\ie\ by looking at the
posterior on the $x$-points of the FK table). In this case we have 
\begin{align}
    \label{eq:bias_function_space_xx}
    \tilde{\mbm} - \mbf_0 = \left[R^{(0)}_{\mbx\mbx} - \mathds{1}\right] \mbf_0\,,
\end{align}
and the covariance of the posterior can be written as
\begin{align}
    \label{eq:cov_decomposition}
    \tilde{K}_{\mbx\mbx} = \left(\mathds{1} - R^{(0)}_{\mbx\mbx}\right) K_{\mbx\mbx} \,.
\end{align} 
Using the FK tables, Eqs.~\eqref{eq:bias_function_space_xx},~\eqref{eq:cov_decomposition}
can be recast in terms of bias $\mathcal{B}$ and variance $\mathcal{V}$ in data space, 
as defined in Ref.~\cite{DelDebbio:2021whr}.
The bias is the difference between the true data $y^{\text{true}}$ and the corresponding theory prediction 
computed using the result of the analysis, and represents the amount by which 
the resulting model fails in reconstructing the true data. 
The variance gives the error of the corresponding theory predictions.
Writing down their explicit expressions Eqs.~\eqref{eq:bias_data_space},~\eqref{eq:var_data_space}, 
it is clear that bias and variance in data space both vanish, 
\begin{align}
    \label{eq:bias_data_space}
    &\mathcal{B}= \FKtab \left(\tilde{\mbm} - \mbf_0\right) 
        = \FKtab \left(R^{(0)}_{\mbx\mbx} - \mathds{1}\right) \mbf_0
        = 0\,,\\
    \label{eq:var_data_space}
    &\mathcal{V} = \FKtab\, \tilde{K}\, \FKtabT 
     = \FKtab \left(\mathds{1} - R^{(0)}_{\mbx\mbx}\right) K_{\mbx\mbx} 
     \FKtabT = 0\,.
\end{align}
In the case of zero experimental error, the methodology reconstructs the input experimental data 
exactly, independently on the specific values of the hyperparameters; 
note that despite the fact that there is no bias in data space, the model
function is not in general reconstructed exactly, \ie\ $\tilde{\mbm} \neq \mbf_0$ (but they are equal if $\FKtab$ has independent columns).
Therefore in the case of infinitely precise data, perfect reconstruction is achieved at the data level, 
but not in the functional space, where a residual reconstruction error is still present.

\begin{figure}[h!]
    \center
    \includegraphics[scale=0.4]{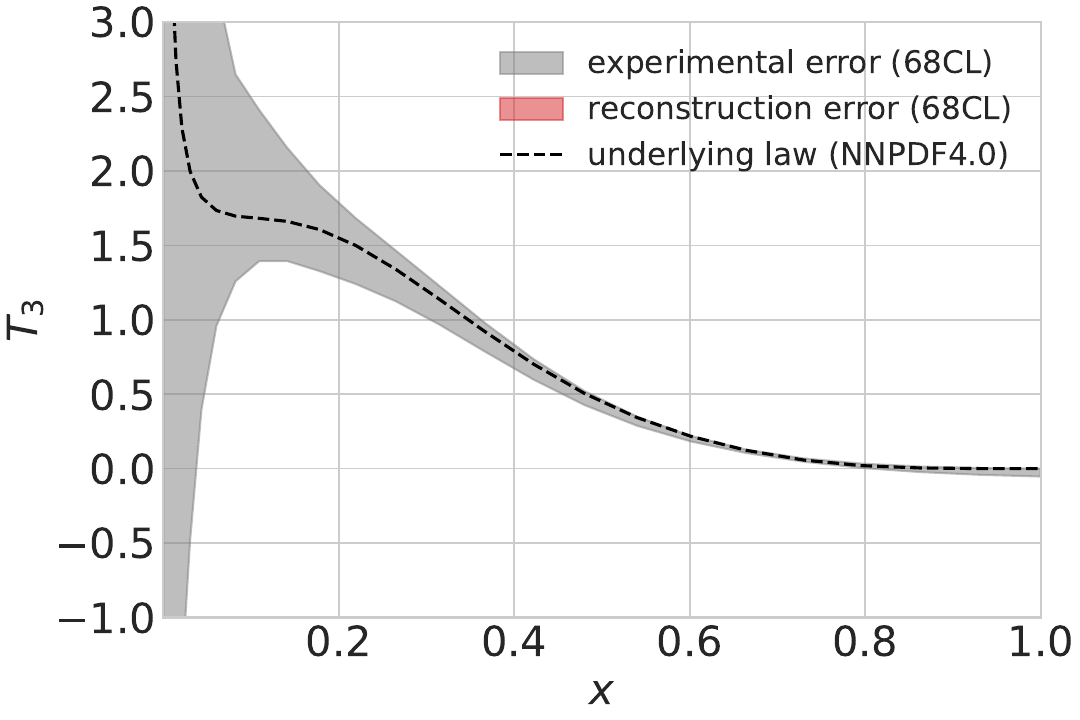}
    \includegraphics[scale=0.4]{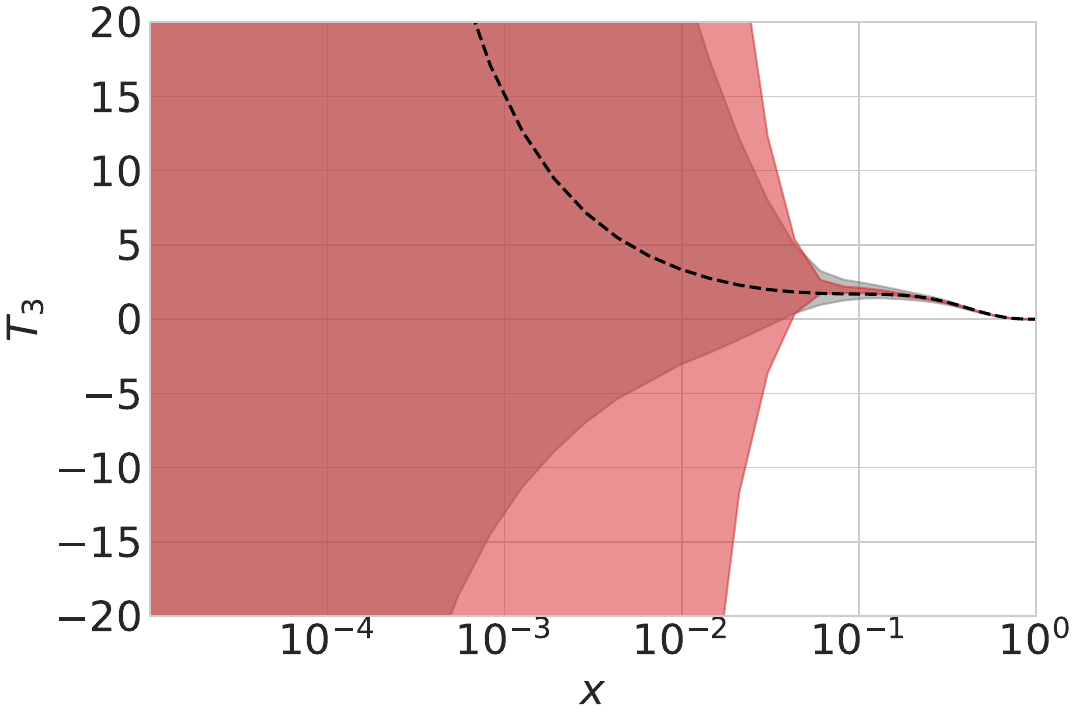}
    \caption{\small Same as Fig.~\ref{fig:mcT3} with the 68\% CL band for 
    the experimental and reconstruction errors plotted separately in grey and red respectively,
    according to Eq.~\eqref{eq:cov_decomposition_cy}. Note: the error bands are overlapped with transparence, rather than stacked.
    \label{fig:exp_meth_cov}
    }
\end{figure}

\paragraph{Non-vanishing experimental error}
Now let us introduce back a non vanishing experimental error.
The reconstruction kernel is
\begin{align}
    \label{eq:ReconstructionKernelWithExpErrors}
    R_{\mbx^*\mbx} = K_{\mbx^*\mbx}\, 
        \FKtabT \left[\FKtab K_{\mbx\mbx} \FKtabT + C_Y\right]^{-1} \FKtab\,,
\end{align}
and Eqs.~\eqref{eq:bias_function_space_xx},~\eqref{eq:cov_decomposition} become
\begin{align}
    \label{eq:bias_function_space_cy}
    &\tilde{\mbm} - \mbf_0 = \left[R_{\mbx\mbx} - \mathds{1}\right] \mbf_0 + 
        a_{\mbx\mbx}^T \eta\, , \\
    \label{eq:cov_decomposition_cy}
    &\tilde{K}_{\mbx\mbx} = \left(\mathds{1} - R_{\mbx\mbx}\right) K_{\mbx\mbx}
        \left(\mathds{1} - R_{\mbx\mbx}\right)^T 
        + a_{\mbx\mbx}^T C_Y a_{\mbx\mbx}\, ,
\end{align} 
where we have introduced
\begin{align}
    \label{eq:aOpDef}
    a^T_{\mbx\mbx} = K_{\mbx\mbx}\,\FKtabT \left[\FKtab \, K_{\mbx\mbx} \,
         \FKtabT + C_Y\right]^{-1}\, ,
\end{align}
so that 
\begin{align}
    \label{eq:aRRelation}
    R_{\mbx\mbx} = a^T_{\mbx\mbx} \FKtab\,.
\end{align}
The corresponding expressions in data space Eqs.~\eqref{eq:bias_data_space},~\eqref{eq:var_data_space}
become
\begin{align}
    \label{eq:bias_data_space_cy}
    &\mathcal{B}= \FKtab \left[R_{\mbx\mbx} - \mathds{1}\right] \mbf_0 + 
    \FKtab a_{\mbx\mbx}^T \mathbf{\eta}\,,\\
    \label{eq:var_data_space_cy}
    &\mathcal{V} = \FKtab \left(\mathds{1} - R_{\mbx\mbx}\right) K_{\mbx\mbx} 
    \left(\mathds{1} - R_{\mbx\mbx}\right)^T \FKtabT
    + \FKtab a_{\mbx\mbx}^T C_Y a_{\mbx\mbx} \FKtabT\,.
\end{align}
Perfect reconstruction is not achieved anymore: bias and variance 
in data space no longer vanish, and their specific value will depend 
on the choice of the hyperparameters for the kernel.
The decomposition in Eqs.~\eqref{eq:bias_function_space_cy},~\eqref{eq:cov_decomposition_cy} 
highlights the fact that
there are two types of contributions to the bias and to the posterior covariance matrix. 
The first term in Eqs.~\eqref{eq:bias_function_space_cy},~\eqref{eq:cov_decomposition_cy} 
comes from the limited reconstruction of the central value and indeed would vanish when
$R_{\mbx\mbx}=\mathds{1}$~\cite{10.1093/gji/ggz520}.
This term survives in the
limit where $C_Y \to 0$, \ie\ in the limit of no experimental errors on the
data, in which case we have $R_{\mbx\mbx} \rightarrow R^{(0)}_{\mbx\mbx}$
and we recover Eqs.~\eqref{eq:bias_function_space_xx},~\eqref{eq:cov_decomposition}. 
The second term is the propagation of the covariance of the data into the
covariance of the model. In the case $R_{\mbx\mbx}=\mathds{1}$, the only error
fluctuations in the posterior distribution come from this term.
In Fig.~\ref{fig:exp_meth_cov}, Montecarlo samples generated 
according to the reconstruction and experimental components are plotted separately for the BCDMS results,
in red and grey respectively.
In the medium-$x$ region, where more experimental data are available, 
the PDF uncertainty is dominated by the experimental error, yet a smaller reconstruction error is still present;
when moving to the small and large-$x$ extrapolation regions the reconstruction error becomes the dominant one, 
pointing out the lack of experimental information. 
We stress once more how these qualitative considerations are precisely quantified 
in Eqs.~\eqref{eq:bias_function_space_cy},~\eqref{eq:cov_decomposition_cy}, giving the analytical expression 
for the posterior covariance matrices associated to the experimental and reconstruction error, 
making it possible to quote different component of the PDF error in the context of a pheno analysis.

\subsection{Validation}
As discussed in the previous section, whenever we deal with noisy experimental information 
perfect reconstruction of the true data is not achieved anymore and different methodologies
are expected to perform differently. In this section we introduce some statistical 
metrics which allow to validate a given methodology.
We discuss the definition of such metrics and what we expect in case of successful reconstruction 
of the underlying law. Finally we compute their probability distribution for the results presented in the previous section. 

\begin{figure}[h!]
    \center
    \includegraphics[scale=0.35]{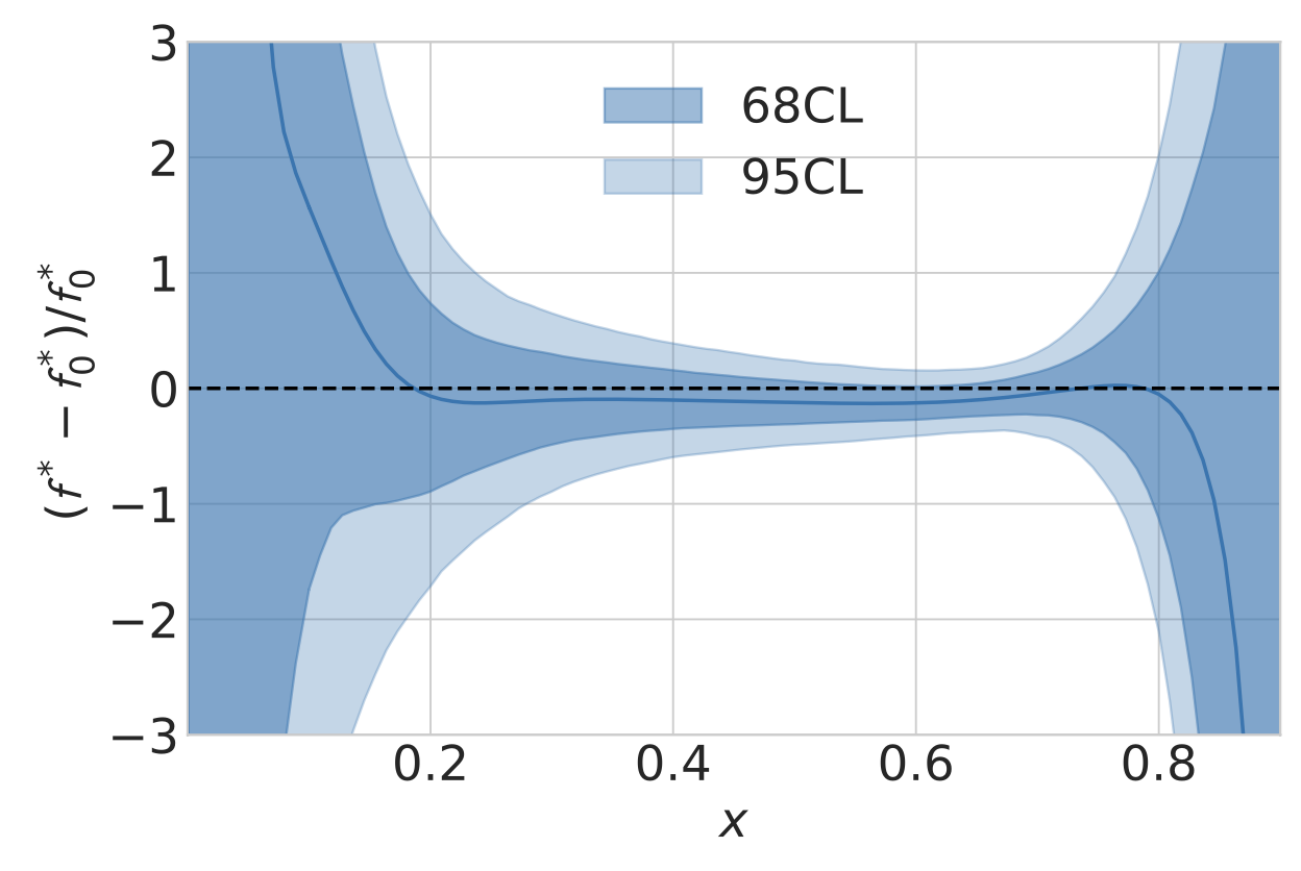}
    \includegraphics[scale=0.38]{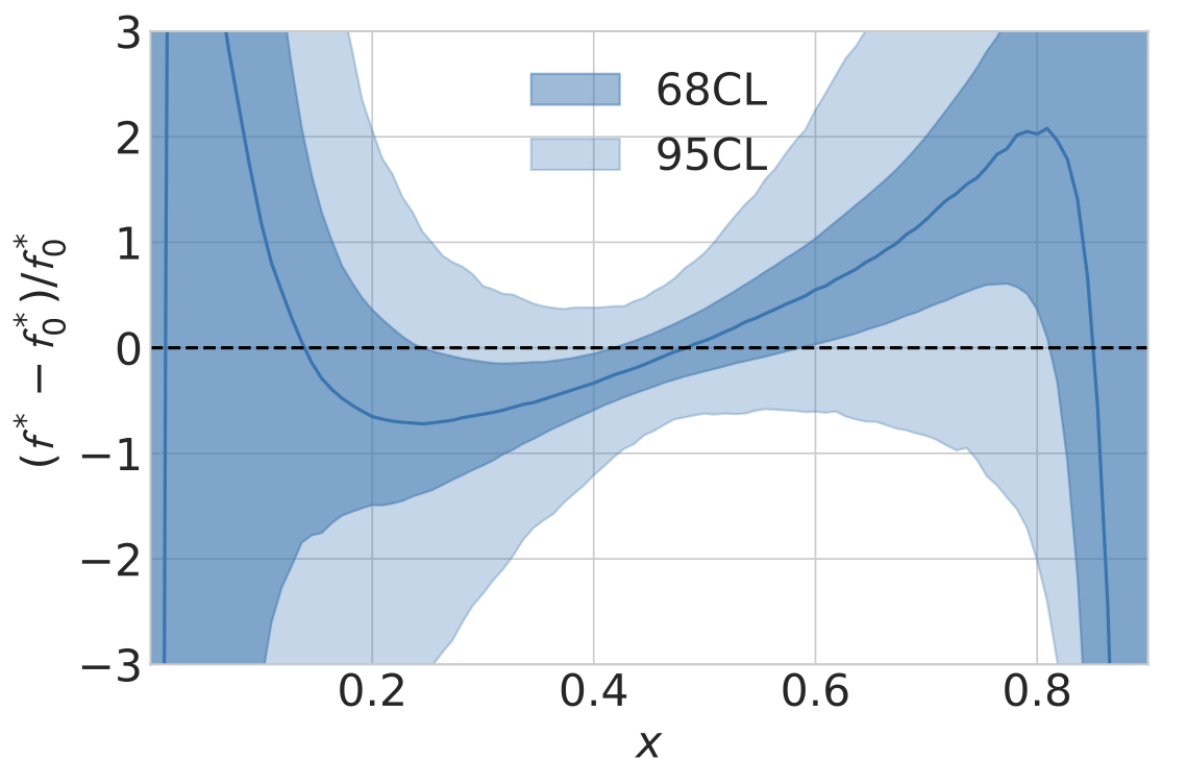}
    \caption{\small Distribution of $\mbf^* -  \mbf_0^* \mid \FKtab\,\mbf + \epsilon = y$ normalized to
    $\mbf^*_0$.
    \label{fig:bias_function_space}
    }
\end{figure}

\paragraph{Closure tests}
In the context of a closure test, \ie\ when the analysis is performed on pseudo-data 
built from a known model, as done in this paper, 
the results can be validated a posteriori, by checking 
how well the posterior distribution describes the underlying law.
A first assessment is obtained by looking at the 
distribution of the stochastic variable
\[
    \mbf^* -  \mbf_0^* \mid \left(\FKtab\,\mbf + \epsilon = y\right)\, ,
\]
whose mean and covariance are given by $\tilde{\mbm}^* -  \mbf_0^*$ and $\tilde{K}_{\mbx^*\mbx^*}$.
In Fig.~\ref{fig:bias_function_space} we plot its distribution,
normalized to $\mbf^*_0$.
The left (respectively, right) panel shows the result for the case of 
BCDMS (respectively, lattice data): the difference is compatible with zero in the full $x$ range, with a 
smaller error in the kinematic region which is more sensitive to the observed data.

In addition to the plot displayed in Fig.~\ref{fig:bias_function_space},
a more quantitative measure of the inference performances can be obtained 
from the value of the log-loss, 
\ie\ the negative logarithm of the posterior probability distribution that the stochastic 
variable $\mbf^*$ is equal to the underlying law $\mbf^*_0$
\begin{align}
    \label{eq:ValidLogLossDef}
    L(\mbf^*_0,\tilde{\mbm^*},\tilde{K}_{\mbx^*\mbx^*}) 
    &= 
        -\log\left(
            P\bigl[\mbf^*  = \mbf^*_0 \mid \FKtab\,\mbf + \epsilon = y\bigr]
        \right)\, , \\
    \label{eq:ValidLogLossEval}
    &=  \frac{n^*}2 \log(2\pi)
        + \frac12 \log\pdet\tilde K_{\mbx^*\mbx^*} 
        + \frac12 (\mbf^*_0 - \tilde\mbm^*)^T \tilde K_{\mbx^*\mbx^*}^+ (\mbf^*_0 - \tilde\mbm^*)\,.
\end{align}
Note that in Eq.~\eqref{eq:ValidLogLossDef}, $\mbf^* \mid \left(\FKtab\,\mbf + \epsilon = y\right)$ is a 
stochastic variable, whose probability density is given by the posterior Gaussian Process, 
while $\mbf^*_0$ is the known underlying law. For probabilities in $(0, 1)$, 
the log-loss is 0 when the posterior assigns probability \SI{100}\% 
to what is actually observed, while with probability densities the scale is arbitrary. 
This means that in the continuous case the log-loss can be used to compare and benchmark inferences, 
but its absolute value does not have a definite interpretation. 
As one can see in Eq.~\eqref{eq:ValidLogLossEval}, the log-loss penalizes not only wrong answers, \ie\ 
posterior central values that differ from the underlying law, 
but also large errors. 

Although here we wrote out the expression of the log-loss for our specific GP model 
(Eq.~\ref{eq:ValidLogLossEval}), the definition of log-loss is totally general within the Bayesian paradigm, 
so it could be used to compare any other fully Bayesian inference to our method. 
We do not try such a comparison in this paper.

\begin{figure}[h!]
    \center
    \includegraphics[scale=0.4]{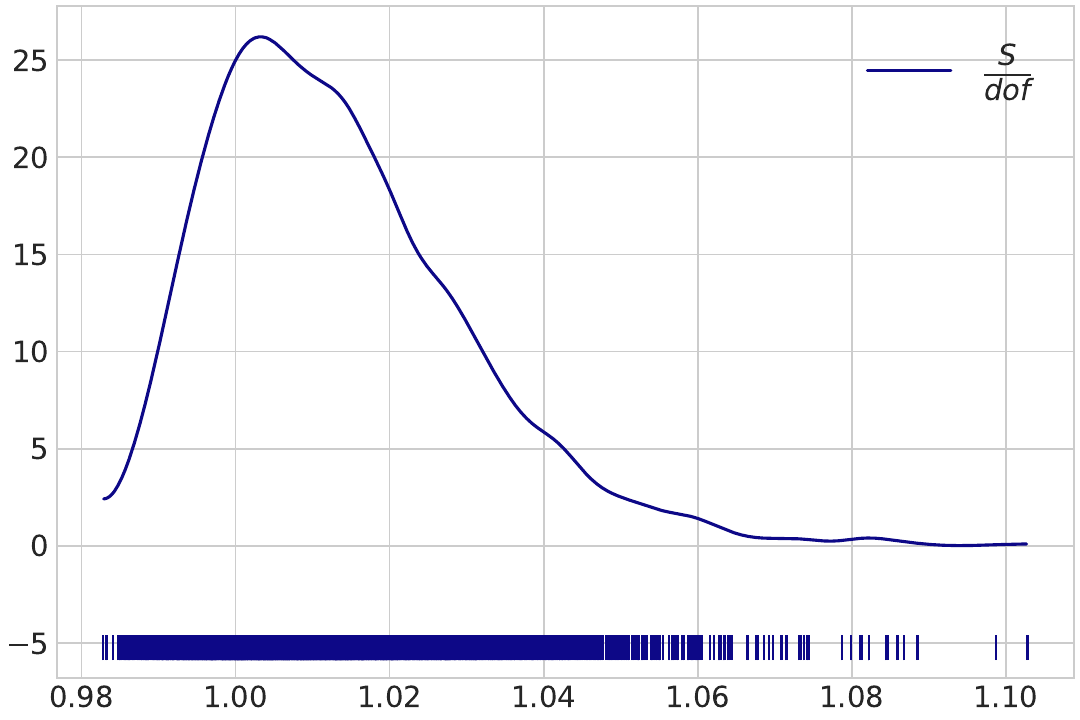}
    \includegraphics[scale=0.4]{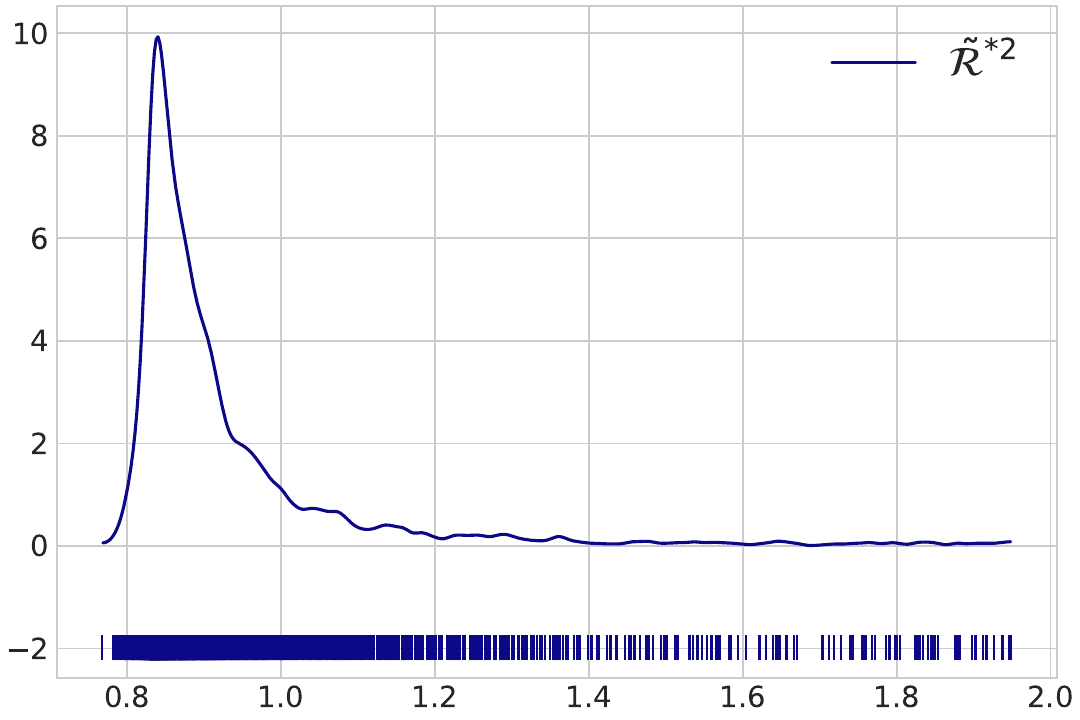}
    \caption{\small $S/\mathrm{dof}$ (left panel) and $\tilde{\mathcal{R}}^{*2}$ (right panel).
    The distributions are obtained by sampling from the hyperparameter posterior 
    $p\left(\theta|y,C_Y\right)$ and using the explicit expressions in Eqs.~\eqref{eq:logloss_validation},
    \eqref{eq:Rtilde_validation}.
    \label{fig:chi2}
    }
\end{figure}

\paragraph{Real data analysis}
When dealing with a real analysis the true underlying model is not known.
Some metrics assessing a posteriori the goodness of the fit and the ability of the model to 
generalize to unseen data are therefore necessary to evaluate the performance of a 
given methodology. A possible metric is given by the quantity
introduced in Eq.~\eqref{eq:LogLossIntroduced} evaluated for $\mbf=\tilde{\mbm}$,
which is in some proper statistical sense the Bayesian equivalent of a more familiar frequentist $\chi^2$:
\begin{align}
    \label{eq:logloss_validation}
    \frac{S(\tilde{\mbm}; \theta, y, C_Y)}{\mathrm{dof}} &=
        \frac{1}{N_\text{data}} \Big(
        (\mbm - \tilde\mbm)^T K_{\mbx\mbx}^{-1}(\mbm - \tilde\mbm)
        + (y - (\fk)\tilde\mbm)^T C_Y^{-1}(y - (\fk)\tilde\mbm) \Big)\,.
\end{align}
The two pieces can be looked at separately to see if the fit is deviating from
the prior or from the data. 
The usual empirical usage of expecting $S/\mathrm{dof} \approx 1$ is valid.
Its distribution is plotted in Fig.~\ref{fig:chi2}, left panel.

When quantifying the performance of the model
on a test set\footnote{As customary, we denote as training data the set of data entering hyperparameter 
inference and gaussian conditioning, and as test data a set of out-of-sample data, \ie\ a
set of data not entering neither hyperparameter inference nor gaussian conditioning.
We can think of test data as a new dataset with its own covariance $C_Y^*$, uncorrelated to the
data used for the Bayesian inference.}
we can compute the log-loss
$$-\log P(\text{test}|\text{data})\,.$$
Denoting with $*$ the quantities computed on the test points this is given by
\begin{align}
    \label{eq:logp2}
    L(y^*) &= \frac{n^*}2 \log(2\pi)
    + \frac12 \log\pdet\big(\FKtabs \tilde K_{\mbx^*\mbx^*} \FKtabs^T + C_Y^* \big) + {} \notag \\
    &\phantom{{}={}} + \frac12 (y^* - \FKtabs\tilde\mbm^*)^T \big(\FKtabs \tilde K_{\mbx^*\mbx^*} \FKtabs^T + C_Y^*\big)^+ (y^* - \FKtabs\tilde\mbm^*).
\end{align}
Excluding the first constant term, the other two terms are interpretable: 
the first is a log-determinant of the posterior covariance matrix, 
so it's a measure of the volume occupied by the distribution; 
in other words, it summarizes how large is the final uncertainty. 
The second term is the usual squared distance between prediction and data in units of the uncertainty.
Having this in mind we can define the metrics
\begin{align}
    \label{eq:Rtilde_validation}
    \tilde{\mathcal{R}}^{*2} &= 
    \frac{1}{\dim(y^*|y)}
    (\FKtabs \tilde\mbm - y^*)^T 
    \left(\FKtabs\tilde{K}_{\mbx\mbx}\FKtabs^T + C_Y^*\right)^+
    (\FKtabs \tilde\mbm - y^*), \\
    \label{eq:sigma_validation}
    \tilde{\sigma}^{*2} &= \exp\left( \frac{1}{\dim(y^*|y)} \log\pdet \left(\FKtabs\tilde{K}_{\mbx\mbx}\FKtabs^T + C_Y^*\right) \right).
\end{align}
The quadratic form $\tilde{\mathcal{R}}^{*2}$ should be about~1 
(to be sure that test data are described within uncertainty), while $\tilde\sigma$ quantifies
the average posterior error in a Bayesianly justified way. 
By comparing the $\tilde\sigma$ values corresponding to different methodologies we can 
assess quantitatively which one is more or less conservative.

In the context of the simple example presented in this paper, we can use the BCDMS 
and lattice data as training and test set respectively.
Using samples from $p\left(\theta|y,C_Y\right)$ and Eqs.~\eqref{eq:Rtilde_validation},
we can access the full probability distribution of $\tilde{\mathcal{R}}^{*2}$, 
which is plotted in Fig.~\ref{fig:chi2}, right panel.

In this paper we do not try to compare our methodology to the standard results obtained 
with a non-Bayesian approach, as it would require a substantial amount of work 
beyond the scope of the rest of the paper. We leave it to a future analysis with a more complete DIS dataset. 
We just comment on how the goodness-of-fit metrics we have shown here would (or wouldn't) apply:
the log-loss is defined only within a Bayesian inference, 
so it would not allow comparisons between our methodology and the standard ones. 
It would only allow comparisons, say, between different choices of kernel for the GP, 
or between a GP and a non-GP but still Bayesian model.
On the other hand, the quantity $\mathcal R^{*2}$ from Eq.~\eqref{eq:Rtilde_validation} 
can be generalized to a standard fit - for example, in the case of the MonteCarlo fits 
carried out within the NNPDF methodology, the posterior mean $\tilde{\mathbf m}$ 
and covariance matrix $\tilde K_{\mathbf x\mathbf x}$ should be replaced by 
the mean and covariance matrix of the replicas -  
and would therefore allow for a quantitative comparison 
between our methodology and a non-Bayesian approach.

%% file: summary.tex
\section{Summary and future work}
\label{sec:conclusions}

We have described a Bayesian methodology for the solution of the inverse problem 
underlying the determination of PDFs.
GPs are used for the modelling of the PDF prior. Known physical constraints,
such as sum rules, kinematic limit and small-$x$ power behaviour are 
implemented in the prior by suitable manipulation of the GP kernel.
We discussed the case of observables that depend linearly on the PDF, and the analytical
simplification occurring in this scenario, and we applied the methodology to two simple
examples concerning the extraction of a single PDF flavor from a reduced dataset of DIS
structure functions and lattice correlators.
In order to validate our approach we have used pseudo-data produced from a known underlying law.
We have found that, even in the presence of noisy data, the input model
is reconstructed within the quoted error. 
We have discussed the mathematical definition of the final uncertainties given by this approach, 
which allows for a quantitative  estimation of the different components entering the PDF error.
Finally we have discussed the validation of the results by introducing a 
set of metrics, which allow to assess the goodness of a given methodology and compare 
different ones, using Bayesianly justified figures of merit.

This work is intended to be a preliminary study to explore the main features, advantages
and limitations of the Bayesian approach, and it paves the way to a full PDF
determination from an extended DIS dataset.
This will be the object of a future separate paper, in which we aim to deliver a full
DIS-only PDF set, to be compared to other available sets based on parametric regression.

In order to achieve a global PDF determination, not only based on DIS data but including
also LHC hadronic observables, the general approach described here can still be applied,
but no analytical simplification occur, as described in Sec.~\ref{sec:QuadCase}.
This implies that a Monte Carlo with dimension given by the total number of parameters
and hyperparameters has to be run to access the full posterior, which makes the
problem computationally more expensive than the linear case, where the Monte
Carlo dimension is given by the number of hyperparameters only. 
The general features of the Bayesian approach still hold, and the development of a
framework for a global PDF determination will be the object of a further studies.

\paragraph{Acknowledgments}
TG is supported by NWO via an ENW-KLEIN-2 project. The work of LDD was supported by the ExaTEPP project EP/X01696X/1,
and by the UK Science and Technology Facility Council (STFC) grant ST/P000630/1.

We thank Juan Rojo and the members of the NNPDF collaboration for comments and discussions.

%% file: prior_pdf.tex
\section{Sum rules}
\label{app:prior_PDF}
Following the discussion in Section~\ref{sec:Gauss}, we associate a GP to each
PDF in the evolution basis. 
The valence distributions $V_{\sfa}$ obey sum rules that we want to incorporate into
the prior. This can be done by associating a GP to the
indefinite integral of the PDF:
denoting as $\widehat{V}_{\sfa}$ the primitive of $V_{\sfa}$, we associate to the former a GP
having mean and kernel 
\begin{align}
    \widehat{m}^{\sfa}\left(x\right) \quad \text{and} \quad \widehat{K}^{\sfa}\left(x,y\right)\,.
\end{align}
It can be shown~\cite{books/lib/RasmussenW06} that $V_{\sfa}$ is then represented 
by a GP having as mean  and kernel
\begin{align}
    \label{eq:kernel_derivative}
    &m^{\sfa}\left(x\right) = \partial_x \widehat{m}^{\sfa}\left(x\right)\,,
    &K^{\sfa}\left(x,y\right) = 
    \partial_x\partial_y \,\widehat{K}^{\sfa}\left(x,y\right)\, .
\end{align}
In formulae
\begin{align}
    \label{eq:VPriors}
    V_{\sfa}(x) = \widehat{V}'_{\sfa}(x)\, , \quad 
    &\widehat{V}_{\sfa} \sim \GP\bigl(0,\widehat{K}^{\sfa}\left(x,y\right)\bigr)\, , \nonumber\\ 
    &V_{\sfa} \sim \GP\bigl(0,K^{\sfa}\left(x,y\right) \bigr)\, , \quad \sfa=1,3,8,15\, ,
\end{align}
where we used $V_1(x)$ to denote the total valence $V(x)$, the ${}'$ denotes the
derivative with respect to $x$, and $K^{\sfa}\left(x,y\right)$ is given in Eq.~\eqref{eq:kernel_derivative}.
The sum rules can then be expressed by introducing additional GPs for the 
primitive of the PDFs, with kernels satisfying Eq.~\eqref{eq:kernel_derivative}, 
and by imposing linear constraints between them.
In the case of the valence sum rules we get
\begin{align}
    \label{eq:SumRulesConstraintsOne}
    &\widehat{V}(1) - \widehat{V}(0) = 
        \widehat{V}_8(1) - \widehat{V}_8(0) = 
        \widehat{V}_{15}(1) - \widehat{V}_{15}(0) = 3\, , \\
    \label{eq:SumRulesConstraintsTwo}
    &\widehat{V}_3(1) - \widehat{V}_3(0) = 1\, .
\end{align}
Similarly the momentum sum rule is written in terms of
the indefinite integral of $x\Sigma$ and $xg$, denoted as $\widehat{x\Sigma}$ and $\widehat{xg}$,
 and can be imposed by introducing the GPs
\begin{equation}
    \label{eq:SingletPriors}
    \widehat{x\Sigma} \sim \GP(0,\Theta^{\Sigma})\, , 
    \quad 
    \widehat{xg} \sim \GP(0,\Theta^{g})\, , 
\end{equation}
and imposing the linear constraint
\begin{equation}
    \label{eq:MomSumRule}
    \widehat{x\Sigma}(1) + \widehat{xg}(1) - 
    \widehat{x\Sigma}(0) - \widehat{xg}(0) = 1\, .
\end{equation}

The power behaviour for $x\to 0^+$ of a given PDF can be enforced by rescaling
the corresponding kernel function according to Eq.~\eqref{eq:smallx_behaviour}. 
In the case in which sum rules and small-$x$ power behaviour need to be imposed at the same time,
the rescaling function of Eq.~\eqref{eq:rescaling_function} 
should be applied at the kernel representing the primitive. 
For example, if we want te valence distribution $V$ to scale as $x^{\alpha_V}$, then 
the kernel for $\widehat{V}$ should be rescaled using
\[
    \phi\left(x\right) = x^{\alpha_V+1}\,.
\]

Since for the primitives we use a Gibbs kernel (Eq.~\eqref{eq:gibbs_kernel})
with variable length scale, the length scale affects the amplitude of the
derivatives. The length scale is proportional to $x$, so the derived process has
standard deviation proportional to $1/x$. In our specific case, it turns out
that the correction does not alter the intended variance of the derived process:
$\partial(\phi\left(x\right) f(x)) \sim x^{\alpha_V} \cdot 1 + x^{\alpha_V + 1}
\cdot x^{-1} \sim x^{\alpha_V}$.

%% file: kinlim.tex
\section{Change of the prior in the extrapolation region: kinetic limit}
\label{app:kinlim}
\begin{figure}[h!]
    \center
    \includegraphics[scale=0.55]{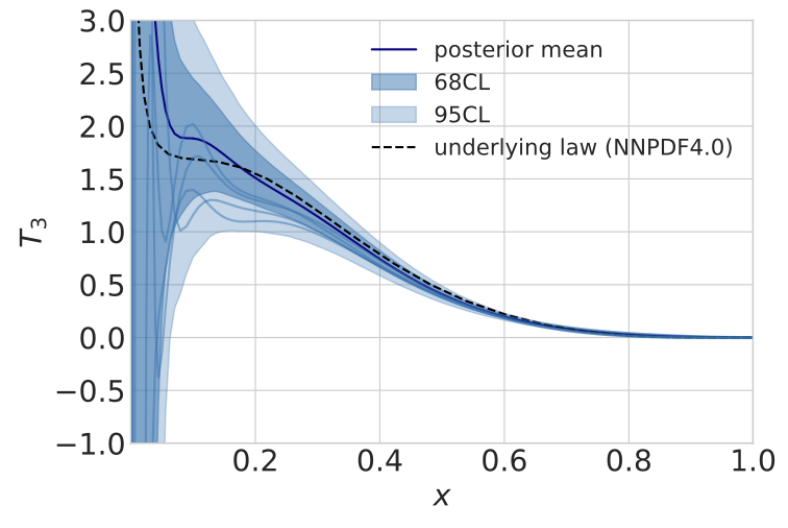}
    \includegraphics[scale=0.55]{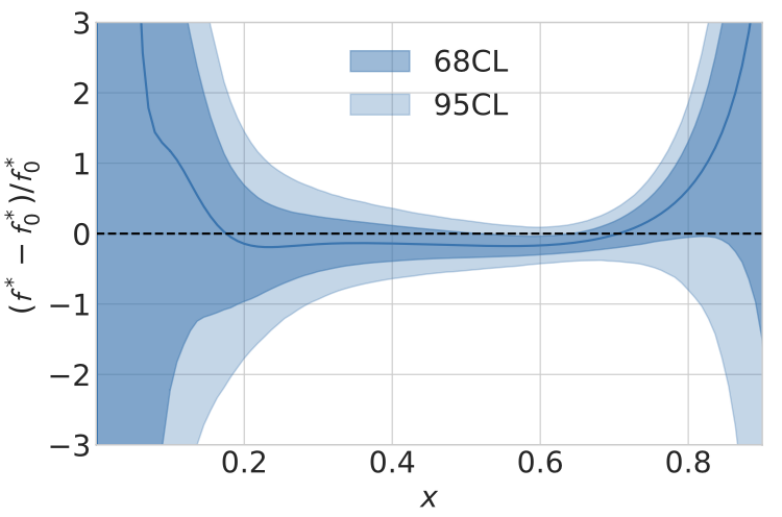}
    \caption{\small Samples from  $p\left(\mbf^*,\alpha,l_0,\sigma | y,C_Y\right)$
    plotted in linear scale (left panel) and
    distribution of $\mbf^* -  \mbf_0^* \mid \FKtab\,\mbf + \epsilon = y$ normalized to
    $\mbf^*_0$ (right panel). Both plots refer to the case of the analysis on the BCDMS 
    data accounting for the kinetic constrain $T_3\left(1\right)=0$.
    \label{fig:kinlim}
    }
\end{figure}

As pointed out in Sec.~\ref{sec:discussion_res}, the behavior of 
the posterior in the extrapolation regions strongly depends on the specific
choice we make for the prior: in the absence of any experimental information 
the posterior reduces to the prior. 
The kinematic constraint according to which all flavors vanish at $x=1$,
known as kinetic limit, is an example of a property that, when implemented, 
directly modifies the prior in the large-$x$ extrapolation region. 
Given that the conditions 
\begin{equation}
    \label{eq:KinCon}
    \Sigma(1) = g(1) = T_{\sfa}(1) = V(1) = V_{\sfa}(1) = 0\, , \quad \sfa=3,8,15\,,
\end{equation}
are simple linear constraints involving each individual flavor,
they can be implemented in the prior by treating them as additional datapoints, extending the FK table.
In the left panel Fig.~\ref{fig:kinlim} we show the results we got for the posterior
when performing the analysis on the BCDMS data accounting for the kinetic limit:
unlike the analogous plot in Fig.~\ref{fig:mcT3} - where at large $x$ the error%
increases reflecting the lack of experimental data - the error in $x=1$ now shrinks to $0$,
according to the new information we implanted in the prior.
In the right panel of Fig.~\ref{fig:kinlim} we plot the distribution of 
\[
    \mbf^* -  \mbf_0^* \mid \left(\FKtab\,\mbf + \epsilon = y\right)\,,
\]
which allows to check that the additional constraint is not introducing a bias.
Comparing this result with the analogous plot in Fig.~\ref{fig:bias_function_space},
it is clear how by imposing the kinetic limit we obtain a better description of
the underlying law at large-$x$.